\documentclass[preprint,5p,twocolumn,11pt,sort&compress]{elsarticle}
\usepackage{graphicx}	% Including figure files
\usepackage{amssymb}% Extra maths symbols
\usepackage{amsmath}	% Advanced maths commands
\usepackage[dvipsnames]{xcolor}    % can be removed once comments are removed
\usepackage[super]{nth}
\usepackage{microtype}
\usepackage{physics}
\usepackage[math]{cellspace} % stretching out the cells in Table 2 a bit to make it easier to read
\usepackage{hyperref}
\hypersetup{
    colorlinks=true, % false: boxed links; true: colored links
    linkcolor=royalblue, % color of internal links (change box color with linkbordercolor)
    citecolor=magenta}
\usepackage{cleveref} % enables \cref, \Cref for referencing multiple figures in one

\def\fabc#1{#1}

\newcommand{\ie}{\emph{i.e. }}

\journal{Physics of the Dark Universe}

\begin{document}

\begin{frontmatter}

\title{Strongly lensed supernovae as a self-sufficient probe of the distance duality relation}

\author[Leiden]{Fabrizio Renzi \corref{cor}}
\author[ICG]{Natalie B. Hogg}
\author[IFT]{Matteo Martinelli}
\author[IFT]{Savvas Nesseris}

\address[Leiden]{Institute Lorentz, Leiden University, PO Box 9506, Leiden 2300 RA, The Netherlands }
\address[ICG]{Institute  of  Cosmology  and  Gravitation,  University  of  Portsmouth, Burnaby  Road,  Portsmouth,  PO1  3FX, United Kingdom}
\address[IFT]{Instituto de F\'isica T\'eorica UAM-CSIC, Campus de Cantoblanco, E-28049 Madrid, Spain}
\cortext[cor]{Corresponding author: renzi@lorentz.leidenuniv.nl}

\begin{abstract}
The observation of strongly lensed Type Ia supernovae enables both the luminosity and angular diameter distance to a source to be measured simultaneously using a single observation. This feature can be used to measure the distance duality parameter $\eta(z)$ without relying on multiple datasets and cosmological assumptions to reconstruct the relation between angular and luminosity distances. In this paper, we show how this can be achieved by future observations of strongly lensed Type Ia systems. Using simulated datasets, we reconstruct the function $\eta(z)$ using both parametric and non-parametric approaches, focusing on Genetic Algorithms and Gaussian processes for the latter. In the parametric approach, we find that in the realistic scenario of $N_{\rm lens}=20$ observed systems, the parameter $\epsilon_0$ used to describe the trend of $\eta(z)$ can be constrained with the precision achieved by current SNIa and BAO surveys, while in the futuristic case ($N_{\rm lens}=1000$) these observations could be competitive with the forecast precision of upcoming LSS and SN surveys. Using the machine learning approaches of Genetic Algorithms and Gaussian processes, we find that both reconstruction methods are generally well able to correctly recover the underlying fiducial model in the mock data, even in the realistic case of $N_{\rm lens}=20$. Both approaches learn effectively from the features of the mock data points, yielding $1\sigma$ constraints that are in excellent agreement with the parameterised results. 
\end{abstract}

\begin{keyword}
Cosmology \sep strong lensing \sep forecasts


\end{keyword}

\end{frontmatter}


\section{Introduction} \label{sec:intro}
The HOLISMOKES project recently demonstrated that the exciting possibility of using strongly lensed Type Ia supernovae (SNIa) as a precision probe in cosmology could soon become a reality \cite{Suyu:2020opl}. Strong gravitational lensing occurs when a massive object lies along the line of sight between a luminous source and an observer. The gravitational field of the lens distorts the spacetime along the line of sight, bending the light path of photons coming from the source which results in a remapping of the source light into multiple images \cite{Grav_lensing1992,Grav_lensing2006}. 

Due to the different light paths taken by photons coming from the source, these images arrive at the observer at different times and are therefore delayed with respect to one another. The time delay between images, which can be measured up to an arbitrary length of time \cite{Borra1997, Borra2008}, is a typical lensing observable which is only sensitive to the mass profile of the lens and to a combination of the source and lens angular diameter distances, the so-called time delay distance \cite{Grav_lensing1992,Grav_lensing2006}.  Provided that one can properly reconstruct the lens mass profile, the strong lensing time delay can then be used as a tracer of the distance--redshift relation, and to infer constraints on cosmological parameters \cite{Treu:2010uj,Treu:2016ljm,Suyu:2018vqs,Shiralilou:2019div}. 

While lensing can happen at all scales (\ie the lens can be as small as a star or as big as a galaxy cluster), for cosmological inference one typically relies on galaxy--galaxy lensing events \cite{Treu:2010uj}. This is because galaxies are believed to have simple mass profiles that can be effectively parameterised as a power law, and a larger lensing probability, making them more abundant in the sky, although it has been shown that uncertainties in the mass profiles play a significant role in constraining cosmological parameters \cite{Birrer:2020tax}. Furthermore, by combining  measurements of the velocity dispersion of the stars orbiting the lensing galaxy with the strong lensing time delay, it is possible to obtain a measurement of the angular diameter distance to the lens, which breaks the degeneracy between different lens mass profiles \cite{Suyu:2009by,Paraficz:2009xj,Jee:2014uxa,Suyu:2016qxx}.

However, strong lensing observables are significantly affected by the specific alignment between the lens and the source, making only a fraction of the observed lensing events suitable for the extraction of cosmological information \cite{Treu:2013rpx,Liao:2014cka}. To achieve percentage accuracy on time delay measurements, the image separation is required to be $> 1^{''}$, the magnitude of the faintest image $m_i < 21$ in the $i$-band, and the lensing galaxy magnitude $m_i < 22$ \cite{Jee:2015yra}.  Therefore, it is clear that, along with good source--lens alignment, one needs sources with a typical brightness comparable to a galaxy to accurately distinguish the lens galaxy from the lensed images. This has led to the use of lensed quasars as the major cosmological probe in the context of lensing, an approach which has been proven by the H0LiCOW collaboration to be highly successful in deriving cosmological constraints  \cite{Wong:2016dpo,Bonvin:2016crt,Tihhonova:2017mym,Birrer:2018vtm,Rusu:2019xrq,Wong:2019kwg,Chen:2019ejq}.
 
There exists another family of astrophysical objects that have luminosities comparable to that of a galaxy: supernova explosions. The concept of using strong lensing of SNIa  as a cosmological probe was pioneered in 1964 by Refsdal \cite{Refsdal1964b}, who showed that the strong lensing time delays can be used to directly measure the Hubble parameter, $H(z)$. However, since lensed supernovae are thought to be far rarer than lensed quasars, the idea of using them for cosmology has long been considered a fruitless endeavour. This changed with the recent observations of two lensed supernova events (the core collapse supernova ``Refsdal'' \cite{Kelly:2014mwa} in 2014 and the Type Ia supernova iPTF16geu \cite{Goobar:2016uuf} in 2016), which reinvigorated the field \cite{Pierel:2019pnr}. As highlighted by HOLISMOKES \cite{Suyu:2020opl}, cosmology with strongly lensed SNIa will soon be possible with surveys like LSST, which is expected to measure around a thousand such events \cite{LSSTscience,Marshall:2017wph,Goldstein:2018bue,Huber:2019ljb}.

As previously mentioned, gravitational lensing remaps the source light from the source plane to the lens plane. While the source surface brightness is conserved in the process, the area on the lens plane in which source photons are remapped is not conserved. In other words the flux of the lensed images is different from the source flux, their ratio defining the magnification factor. From lensing observations, one typically measures the ratio of magnification between the images by comparing their measured fluxes, but the total magnification is not directly measurable because the unlensed source brightness (\ie the unlensed source flux) is unknown. So, despite their relative rarity in comparison to lensed quasars, lensed SNIa have one compelling advantage: they allow the source brightness to be measured independently from lensing observations \cite{Oguri:2002ku}.

By assuming that SNIa are standardisable candles, the brightness (and brightness decay after the explosion) can be inferred from the light curves of the lensed events, which are well known from unlensed supernovae observations. The total magnification can then be tightly constrained, reducing the uncertainties in the lens mass profile and improving the possible cosmological constraints \cite{Oguri:2002ku}. Since this enables us to measure the luminosity distance to these events, they can be used to test more fundamental aspects of the standard cosmological model.

We note that microlensing and other lensing effects related to substructures (such as dust clouds and subhalos) in the deflector galaxy can significantly affect the standardisable nature of SNIa, leading to large uncertainties in the inferred unlensed flux \cite{Yahalomi:2017ihe,Foxley-Marrable:2018dzu,Oguri:2002ku,Goldstein:2018bue,Huber:2019ljb}.  
However, it is expected that a significant fraction of lensed SNIa will be standardisable: around $20\% $ from an LSST-like survey \cite{Foxley-Marrable:2018dzu,Bonvin:2018lgh,Huber:2019ljb}.
In the following, we assume the effect of microlensing and other effects related to substructures in the lensing galaxy to be negligible.

The distance duality relation (DDR), which relates luminosity distances to angular diameter distances, is one example of a fundamental component of cosmology which is accessible with strongly lensed SNIa. Combining information from the velocity dispersion of stars in the lensing galaxy with lensing observations and supernova light curves, lensed SNIa can provide both measurements of angular diameter and luminosity distance, making these events particularly well-suited to probing the DDR and investigating any possible deviations from it, which could indicate the presence of new physics.

In this paper, we aim to reconstruct a function related to the DDR using mock datasets of strongly lensed SNIa. We create the mock datasets for an LSST-like survey, testing three cases: realistic (20 useful lensed SNIa as expected by LSST after 10 years of observations \cite{Suyu:2020opl}), optimistic (100 lenses corresponding to the total number of spatially-resolved lensed SNIa by LSST \cite{LSSTscience}) and futuristic (1000 lenses representing the number of events we expect to observe in the next few decades). Using both parametric and non-parametric approaches for our reconstructions, we investigate whether violations of the distance duality relation could be detected with datasets of this size, finding that the realistic LSST-like survey would be competitive with other more traditional probes of the DDR such as the combination of SNIa and BAO observations. 

We note that a similar analysis, involving strong lensing in the context of constraining the DDR, was performed in \cite{Holanda:2015zpz, Holanda:2016msr, Rana:2017sfr}. However, our approach in this paper differs significantly to those previous works. 
The main difference is that in those works it was shown that it is possible to obtain angular diameter distance measurements from strong lensing events in place of other observations able to provide this quantity (such as BAO), but additional distance luminosity measurements were still needed to constrain the DDR.Instead, we show that both the luminosity and angular diameter distances can be measured from strongly lensed SNIa, exploiting the standardisable nature of supernovae explosions along with the ``standard ruler'' nature of strong lensing events. This makes strongly lensed SNIa a self-sufficient probe of the DDR. 
%This is in contrast to previous works in which separate datasets for the distance measures were required.

The structure of our paper is as follows: in \autoref{sec:theory} we present some theoretical aspects of the distance duality relation, in \autoref{sec:lensedsnia} we discuss the physics of the strongly lensed supernovae and the details of the mock data, while in \autoref{sec:analysis} we present our methodology, with the parameterised and non-parametric approaches, and our results. Finally, in \autoref{sec:conclusions} we summarize our conclusions.

\section{The distance duality relation\label{sec:theory}}
The distance duality relation  is given by \cite{Etherington}
\begin{equation}
    d_L(z) = (1+z)^2 d_A(z), \label{eq:ddr}
\end{equation}
where $d_L(z)$ is the luminosity distance and $d_A(z)$ is the angular diameter distance. It holds under the conditions that photons travel along null geodesics in an pseudo-Riemannian spacetime, and that the number of photons is conserved \cite{Ellis2007}. 

The first condition is a fundamental statement about the geometry of spacetime and the photon mass and is only violated in theories of gravity with a non-Riemannian geometry, or in theories where photons do not propagate on null geodesics due to coupling with other fields (see e.g. \cite{Hehl:1976kj, Hehl:1994ue, Hammond_2002, Gabrielli:2006im, Santana:2017zvy}). It is easier to imagine deviations from DDR occurring due to non-conservation of the photon number, for example by absorption or scattering by dust as they propagate to the observer, or via more exotic mechanisms such as the conversion of photons to axions as they interact with cosmic magnetic fields \cite{Bassett:2003vu}. 

In order to investigate these possible deviations from DDR, a function $\eta(z)$ can be defined from \autoref{eq:ddr} as
\begin{equation}\label{eq:etadef}
    \eta(z) = \frac{d_L(z)}{(1+z)^2 d_A(z)},
\end{equation}
which is equal to unity if the DDR is not altered. DDR violation mechanisms are integrated effects, where  photons interact with intervening components along the line of sight. Thus, one can expect that for a photon at redshift zero, such an effect does not have time to take place and no violation of the relation is present, meaning that $\eta(z=0)=1$. This is also clear from \autoref{eq:etadef}, whose limit for $z=0$ is $\lim_{z\rightarrow0}\eta(z)=1$. For this reason, we impose that $\eta(z)$ is equal to $1$ at vanishing redshifts, for both our parametric and non-parametric reconstructions.

The function $\eta(z)$ is also commonly parameterised in the literature (e.g. \cite{Avgoustidis2009, Avgoustidis2010}) as 
\begin{equation}\label{eq:etaeps}
    \eta(z) = (1+z)^{\epsilon(z)},
\end{equation}
where $\epsilon(z) \neq 0$ is equivalent to $\eta(z)\neq 1$, thus indicating a deviation from the standard DDR. To probe this relation and search for violations of DDR, objects for which both a luminosity distance and angular diameter distance are available are needed. This motivates the use of strongly lensed SNIa, which amply fulfil these criteria.

\section{Strongly lensed supernovae\label{sec:lensedsnia}}
A survey of strongly lensed SNIa will observe the distance modulus of the supernovae, \ie the difference between its apparent and absolute magnitude, which is given by
\begin{equation}\label{eq:modulus}
    \mu(z_s) = 5 \log_{10} \left(\frac{d_L(z_s)}{\textrm{Mpc}}\right) + 25,
\end{equation}
and the time delay distance (see e.g. \cite{Suyu:2018vqs}),
\begin{align} \label{eq:timedelaydistance}
    d_{\Delta t} (z_l) = & (1+z_l)(1+z_s) d_A(z_l) d_A(z_s) \nonumber\\
    &\times\left[(1+z_s)d_A(z_s) - (1+z_l)d_A(z_l)\right]^{-1},
\end{align}
where $z_s$ is the redshift of the source and $z_l$ the redshift of the lens. Notice that \autoref{eq:timedelaydistance} only holds under the assumption of flat space, i.e. $\Omega_k=0$, in the context of a flat Friedmann--Lema\^{i}tre--Robertson--Walker metric. In curved space, the second term on the right hand side would become $d_A(z_s,z_l)$. In this paper we want to obtain measurements of $d_A(z_s)$ and therefore the assumption of a flat Universe allows us to isolate this term in the time delay distance expression. We leave the investigation of more general cases for future work. Under this assumption we can invert \autoref{eq:timedelaydistance} and obtain $d_A(z_s)$, and we can write our parameterisation of the distance duality relation in terms of the distance modulus, the angular diameter distance at the lens and the time delay distance as 
\begin{equation}\label{eq:etaTD} 
    \eta(z_s) = \frac{10^{-5+\mu(z_s)/5}}{(1+z_l)(1+z_s)}\left[\frac{1}{d_A(z_l)}-\frac{1+z_l}{d_{\Delta t}}\right](\rm Mpc).
\end{equation}

The number of currently detected lensed SNIa is insufficient for any precise cosmological application, so we turn to mock datasets to forecast our future ability to probe the distance duality relation with these events.

\subsection{Mock dataset}\label{sec:mocks}

To generate our mock datasets, we focus on lensed SNIa for which measurements of the kinematics of the lens galaxy are available, along with time delay observations. In this scenario, strong lensing will provide two independent distance measures at the same time \cite{Suyu:2009by,Jee:2014uxa,Jee:2015yra}: $d_{\Delta t}(z_l)$ and $d_A(z_l)$.  The measurements of the time delay distance of a lens are obtained by combining the observation of time delays between the light curves of multiple images, a lens mass model for the lensing galaxy and a reconstruction of the mass environment along the line of sight \cite{Wong:2016dpo,Bonvin:2016crt,Tihhonova:2017mym,Birrer:2018vtm,Rusu:2019xrq,Wong:2019kwg,Chen:2019ejq}. We therefore consider only these contributions to the uncertainties of $d_{\Delta t}$.

As in \cite{Suyu:2020opl}, to estimate the precision on $d_{\Delta t}$ we conservatively adopt a $5\%$ uncertainty for the time delay and a $3\%$ uncertainty for both the mass profile and the lens environment. Summing these in quadrature we obtain a cumulative uncertainty on $d_{\Delta t}$ of $6.6\%$, in agreement with current constraints from lensed quasars\footnote{The assumed uncertainties correspond to having a perfect knowledge of the lens mass profile and its environment. As detailed in \cite{Birrer:2020tax}, a hierarchical analysis of the lensing observables may lead to higher uncertainties in the time delay distance.} \cite{Chen:2019ejq}.
For the angular diameter distance to the lens, $d_A(z_l)$, we assume a scenario where spatially-resolved observations of the kinematics of the lens galaxy are available, so that the uncertainties of $d_A$ are essentially dominated by the time delay uncertainties. These measurements are expected to be obtained easily after all the SNIa images have faded. We therefore adopt a $5\%$ precision for $d_A$.

\fabc{The missing ingredient of our mock dataset is now the distance modulus $\mu(z_s)$ of the lensed SNIa. This quantity must be reconstructed  starting from the lensed distance modulus of four lensed images. For standardisable candles this implies fitting the lensed light curves, with exactly the same procedure used for unlensed SNIa, to provide an estimate of the lensed distance modulust $\hat{\mu}$ without any cosmological assumption or knowledge of the lens model.  }
\fabc{The unlensed distance modulus is then related to the lensed one by the following relation:
\begin{equation}
    \mu = \hat{\mu} + \frac{5}{2}\log_{10}A
\end{equation}
where $A$ is the magnification factor of the lensed event, defined as the ratio of the lensed to unlensed flux, \ie
\begin{equation}\label{eq.magnification}
    2.5\log_{10} A = 2.5 \log_{10}\left(\frac{f_{\rm lensed}}{f_{\rm unlensed}}\right) 
\end{equation}}
\fabc{This delensing procedure to infer the unlensed distance modulus can be summarised in two simple steps:
\begin{enumerate}
    \item Estimate the lensed magnitude, $\hat{\mu}$, from the observed light curves of the lensed SNIa.
    \item Assume a mass profile to estimate the lensing magnification\footnote{As the unlensed flux is not measured in lensing observations, lensing magnification has to be determined from the lens mass profile. However, the same mass profile is needed to infer the angular and time delay distances and can be found by studying the lens galaxy and its environment \cite{Suyu:2016qxx,Wong:2016dpo,Tihhonova:2017mym,Rusu:2019xrq,Chen:2019ejq}. Another possibility is to get the unlensed magnitude from an external catalogue of unlensed SNIa and estimate the magnification from Eq. \eqref{eq.magnification} \cite{Goobar:2016uuf}. In this case one can still estimate the distance modulus but it would be the same as the one being inferred from the actual SNIa catalogue, spoiling the information of the lensed event except for the redshift $z_s$}, delens the SNIa and obtain the unlensed modulus distance, $\mu$.
\end{enumerate}}

\fabc{Assuming this approach to be feasible for all the systems in our catalogues to infer the unlensed $\mu(z_s)$, we model its error budget due to the SNIa brightness uncertainties following \cite{Astier:2014swa} and to this we add in quadrature the magnification uncertainty:}%2010ApJ...709.1420G,}:
\begin{equation}
    \sigma[\mu(z_s)]^2=\delta \mu(z_s)^2+\sigma^2_{\textrm{flux}}+\sigma^2_{\textrm{scat}}+\sigma^2_{\textrm{intr}} + \frac{25\sigma^2_{\log A}}{4}
\end{equation}
\fabc{where the systematic uncertainties due to flux calibration are given by $\sigma_{\rm flux} = 0.01$, the intrinsic scatter of SNe at fixed colour, also known as colour smearing, is given by $\sigma_{\rm scat} = 0.025$, the intrinsic distance scatter is $\sigma_{\rm intr} = 0.12$ and finally, we also include an irreducible distance modulus error, which we assume affects all events coherently and varies linearly with redshift in the form $\delta\mu(z_s) = e_M z_s$ with $e_M$ drawn from a normal distribution $\mathcal{N}(0,0.01)$ \cite{Astier:2014swa}. %For the error on SNIa absolute calibration, we assume $\sigma_{M_B}, \sim \sigma_{\log H_0} \sim \sigma_{\log D_{\Delta t}}/\sqrt{N_{\rm lens}}$, corresponding roughly to the uncertainty of inferring the Hubble constant from time delay observations in a flat cosmological model from a catalog of $N_{\rm lens}$ lensed events with a quasi model-independent approach  \cite{Collett:2019hrr}\footnote{\fabc{In \cite{Collett:2019hrr}, a catalogue of unlensed SNIa was combined with the H0LiCOW lenses to estimate the Hubble speed but the same approach can be used for a catalogue of lensed SNIa.}}. This is a key advantage of using lensed SNIa instead of a pair of independent catalogues, as the Hubble constant can be determined from time delay measurements and by assuming a specific mass profile. This can therefore be used to anchor the SNIa distance--redshift relation, incidentally reducing the bias (in distance measures) that may come from using catalogues of independent observations (see e.g. \cite{Renzi:2020fnx}). This dependency on the value of the Hubble constant is then removed when considering the DDR since this is a dimensionless quantity. 
For the error on the lensing magnification we assume a $\sim 20\% $ fractional uncertainties, \ie $\sigma_{\log A} = 0.09$ \cite{Oguri:2002ku}.}

To generate the mock, we assume the lens distribution to be uniform in the range $0.1 \leq z \leq 0.9$ and the source redshift to be twice the lens redshift \ie $z_s = 2z_l$ for simplicity. Even though there will be a distribution for the redshifts of the sources this has a small impact on cosmological inference \cite{Coe_2009,Linder:2011dr}.

Assuming a $\Lambda$CDM fiducial cosmology with $H_0 = 70$ km s$^{-1}$ Mpc$^{-1}$ and $\Omega_m = 0.3$ (with $\Omega_k=0$), we calculate the angular diameter distance $d_A(z)$ at the given $z_l$ and $z_s$. From this we can obtain $d_{\Delta t}(z)$ using \autoref{eq:timedelaydistance}, while to compute the fiducial distance modulus $\mu(z)$ we use \autoref{eq:modulus}, obtaining the luminosity distance from $d_A(z)$ through \autoref{eq:etadef}, which implies choosing a fiducial $\eta(z)$. We rely on the parameterised expression of $\eta(z)$ of \autoref{eq:etaeps}, and we choose for our fiducial a constant $\epsilon(z)=\epsilon_0$. We focus on three different choices for this parameter, in order to be able to test the precision of future observations in different scenarios. We choose the standard DDR value $\epsilon_0=0$, and two fiducials with different degrees of departure from DDR, with $\epsilon_0=0.01,\ 0.05$.

Once the fiducial trends for our observables are computed, we obtain the mock datasets by drawing a random Gaussian shift around the fiducial, using the estimated $1\sigma$ uncertainties for $d_A(z_l)$, $d_{\Delta t}(z_l)$ and $\mu(z_s)$:
\begin{align}\label{eq: genmock}
    D_{i, \rm mock} = D_{\rm mock}(z_i) = D_{\rm true}(z_i) + \delta D(z_i),
\end{align}
with $i=1\dots N_{\rm lens}$, $D_{\rm true}$ representing the fiducial value of either $d_A$, $d_{\Delta t}$ and $\mu$,  and $\delta D$ being the corresponding Gaussian deviate. From this we get our mock distances as $D_{i, \rm mock} \pm \sigma_{D(z_i)}$ where $\sigma_{D(z_i)}$ are the 1$\sigma$ uncertainties of the distance considered.
Finally we use \autoref{eq:etaTD} to obtain a mock catalogue for $\eta(z_i)$ from the mock datasets of $d_A(z_l)$, $d_{\Delta t}(z_l)$ and $\mu(z_s)$. To obtain the error on each of the data points of the mock of $\eta(z_i)$, we employ an MCMC-like approach, detailed as follows:
\begin{enumerate}
    \item We construct the distribution of each of the $D_{i,\rm mock}$ distances at each redshift $z_i$ of the catalogue, drawing 10,000 random samples from the assumed distribution for $D_{i,\rm mock}$.
    
    \item We combine each of the 10,000 random samples using \autoref{eq:etaTD} to obtain 10,000 realisations of the distribution of $\eta(z_i)$ at each redshift $z_i$.
    \item We calculate the mean and standard deviation of $\log_{10}\eta(z_i)$ from the $\eta(z_i)$ distributions at each redshift to construct our final mock datasets.   
\end{enumerate}
A more detailed explanation of the procedure followed to construct the mock datasets can be found in \ref{sec:appendixA}.

Our choice to construct the catalogue using $\log_{10}\eta(z_i)$ is motivated by the fact that the distribution of $\eta(z_i)$ are almost log-normal and therefore $\log_{10}\eta(z_i)$ is almost Gaussian distributed around zero \ie $\log_{10}\eta(z_i) \approx \mathcal{N}(0,\sigma_{\log_{10}\eta(z_i)})$. This allows us to derive constraints from our mock catalogues by employing an MCMC approach with a Gaussian likelihood of the form:
\begin{equation}
    -2\ln\mathcal{L} = \sum_{i=1}^{N_{\rm lens}} \frac{\left[\log_{10}\eta(z_i) - \log_{10}\eta^{\rm th}(z_i)\right]^2}{\sigma^2_{\log_{10}\eta(z_i)}} 
\end{equation}
where $\log_{10}\eta^{\rm th}(z_i)$ is the theoretical value of $\log_{10}\eta(z_i)$.

Furthermore, the choice of constructing the catalogue for $\log_{10}\eta(z)$ is also useful for the application of Gaussian processes that we describe in \autoref{subsec:gp} below; this approach requires the choice of a mean prior for the reconstructed function, which is usually assumed to be zero in standard applications. The choice of reconstructing $\log_{10}\eta(z)$ allows us to keep this assumption without significantly biasing the results. 
%Gaussian processes assume data to be Gaussian distributed with zero mean, so it is generally necessary to rescale the data in order to obtain robust predictions using this kind of machine learning technique. This means that constructing our fiducial datasets by taking the base-10 logarithm of the DDR parameter has the advantage of preparing our mocks for the numerical analysis that we perform later in this work. 

\section{Methodology and results \label{sec:analysis}}
In this section we describe the methodology we use in our analysis and our corresponding results. We first use a simple parameterisation of the DDR violation function $\eta(z)$, forecasting the constraints that can be achieved with realistic ($N_{\rm lens}=20$), optimistic ($N_{\rm lens}=100$) and futuristic ($N_{\rm lens}=1000$) mock datasets. We then focus only on the realistic and optimistic datasets and we apply machine learning approaches, namely Genetic Algorithms (GA) and Gaussian processes (GP), to reconstruct $\eta(z)$.

\subsection{Parameterised approach}

We first adopt a simple parameterised approach to forecast the constraints achievable on DDR violation with future strongly lensed SNIa observations. We use the parameterisation of \autoref{eq:etaeps}, and we assume the function $\epsilon(z)$ to be constant, with its value $\epsilon_0$ the free parameter that we want to constrain with our mock dataset.

We build a likelihood module interfaced with the publicly available MCMC sampler \texttt{Cobaya} \citep{Torrado:2020dgo} which compares the prediction for
\begin{equation}\label{eq.etath}
    \log_{10}{\eta^{\rm th}(z)} = {\epsilon_0}\log_{10}{(1+z)}\, ,
\end{equation}
with the mock dataset we described in \autoref{sec:mocks}. 

The improvement brought by strongly lensed SNIa observations to this analysis is evident. In most previous constraints of DDR violations, predictions of both $d_L(z)$ and $d_A(z)$,  which enter in the definition of $\eta(z)$ in \autoref{eq:etadef}, were required, as the two observables are compared independently with data (see e.g. \cite{Avgoustidis:2010ju,Martinelli:2020hud,Hogg:2020ktc}). Such an approach is  intrinsically dependent on the assumptions made about the expansion history of the Universe, and in particular on the assumed dark energy model driving the late time accelerated expansion. Here, such an assumption is not necessary, as the distances entering \autoref{eq:etaTD} are obtained at each redshift from a single observation, and therefore there is no need to assume a cosmological model to reconstruct the luminosity and angular distances.

However, it is important to note that we assume that $\eta(z)$ as defined in \autoref{eq:etadef} is a valid description of DDR violation, which implies that the Universe is to first approximation homogeneous and isotropic. Finally, for \autoref{eq:etaTD} to hold, we further assume that the contributions to the total energy density by curvature are negligible ($\Omega_k=0$).

\begin{figure}[!t]
    \centering
    \includegraphics[width=0.4\textwidth]{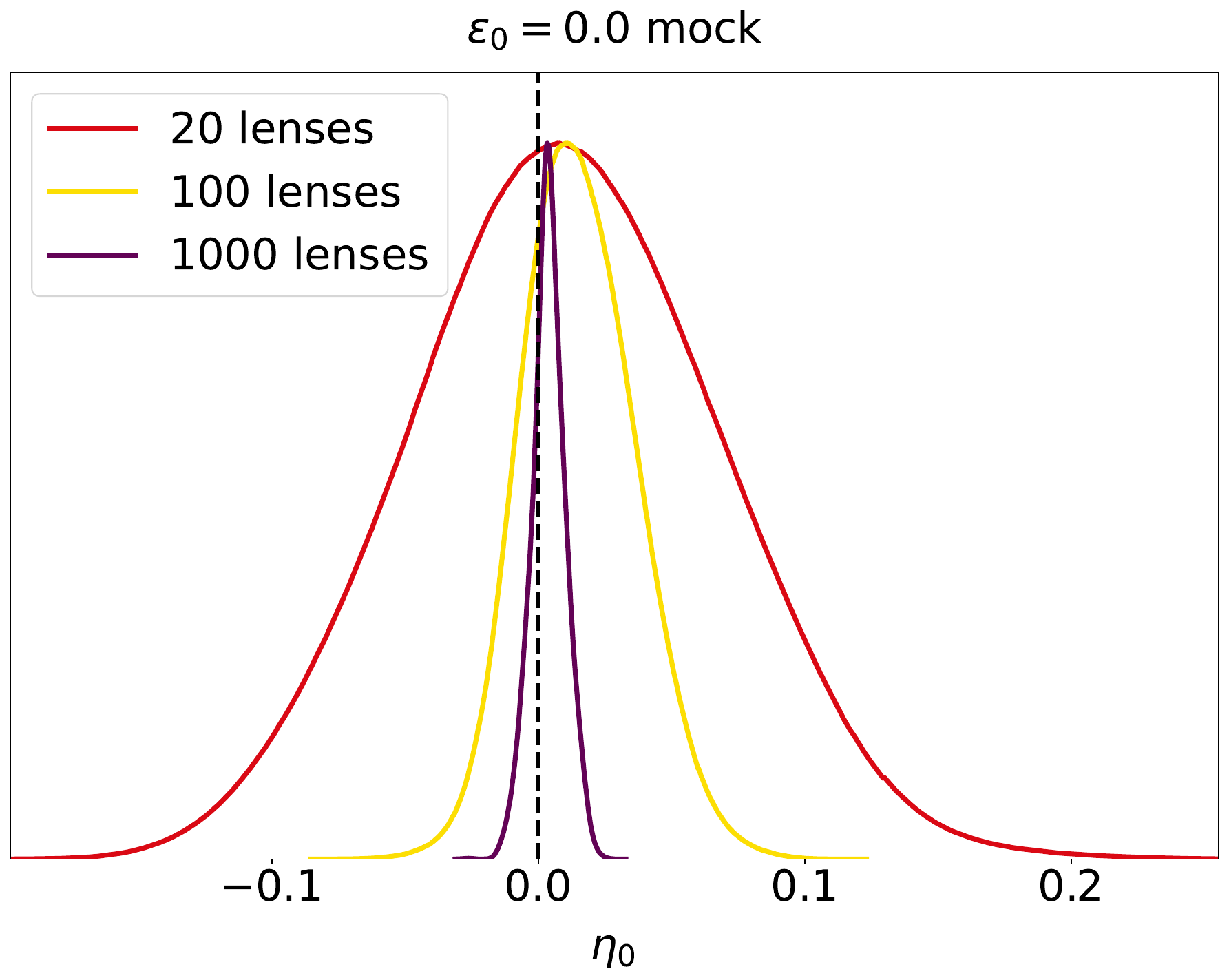}\\
    \includegraphics[width=0.4\textwidth]{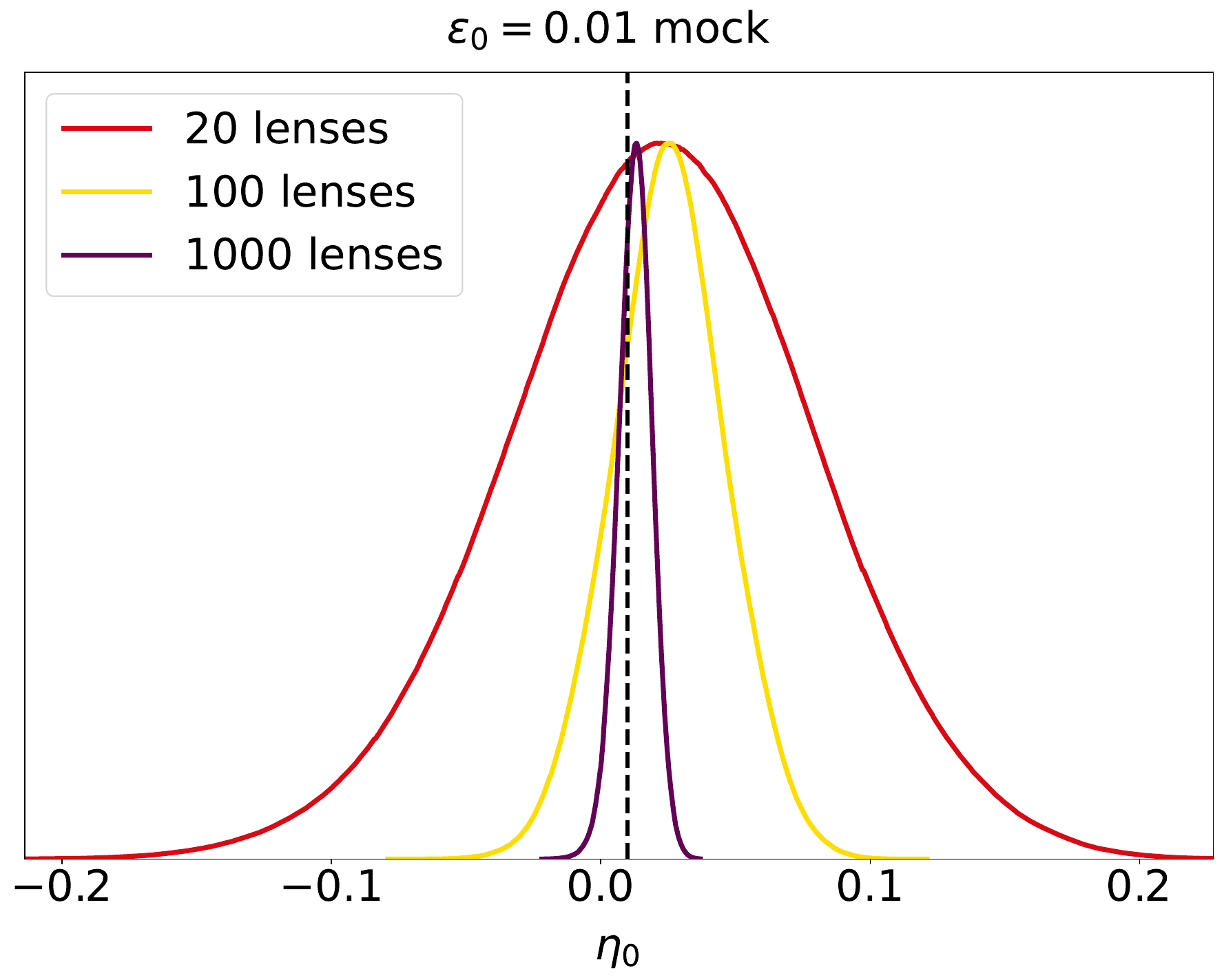}\\
    \includegraphics[width=0.4\textwidth]{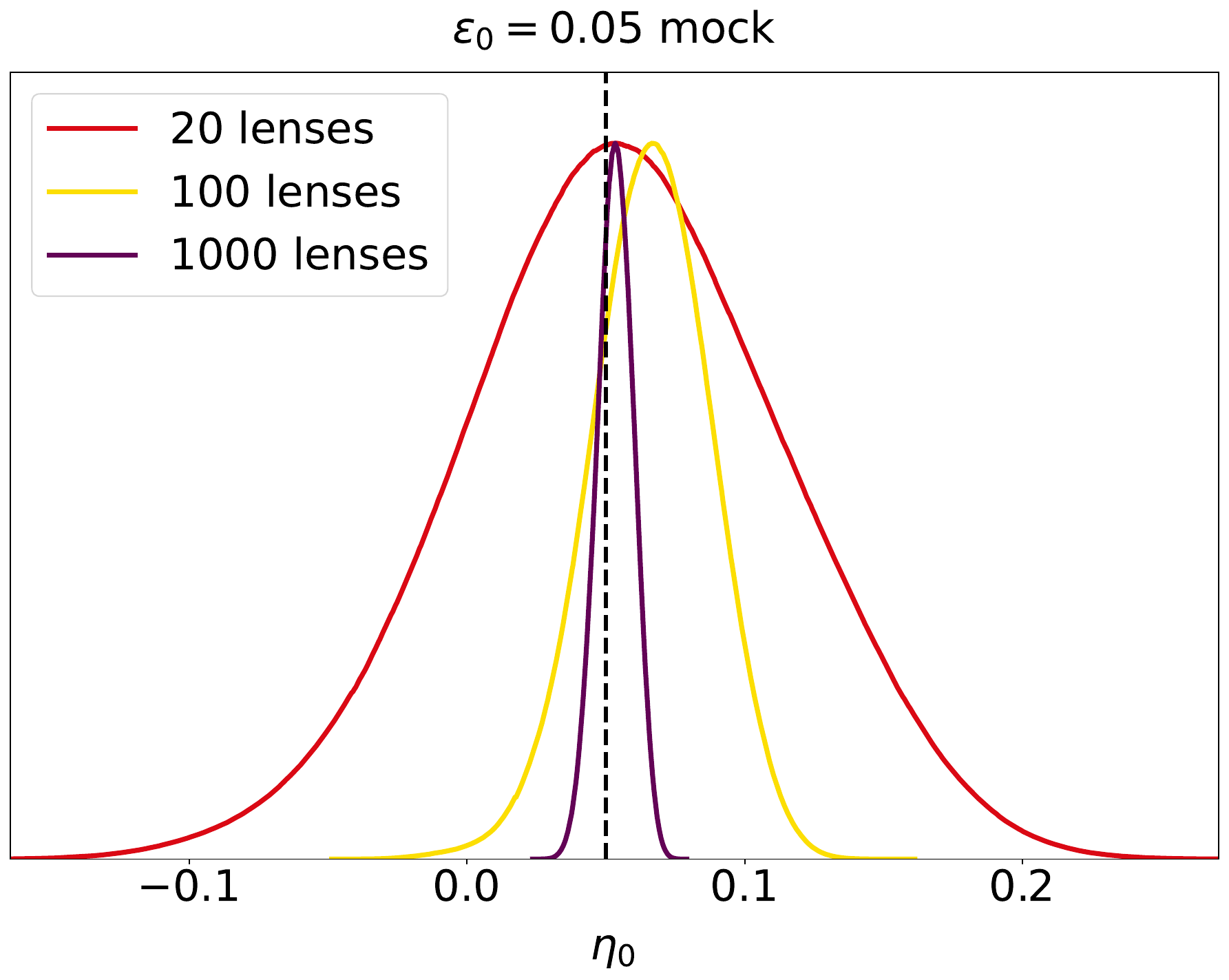}\\
    \caption{Posterior distributions for the DDR violation parameter $\epsilon_0$. The three panels refer to the different fiducial values considered to build the mock data: $\epsilon_0=0$ (top), $\epsilon_0=0.01$ (centre) and $\epsilon_0=0.05$ (bottom). In all panels the different lines show the posterior distribution for the realistic (red), optimistic (yellow) and futuristic (purple) cases.}
    \label{fig:parres}
\end{figure}

For these reasons, the only free parameter in this analysis is $\epsilon_0$, for which we use a flat prior. The constraints we obtain on this are shown in \autoref{tab:parres} and the posterior distributions in \autoref{fig:parres}. We find that the realistic case ($N_{\rm lens}=20$) would achieve the same constraining power of current constraints obtained through the combination of SNIa and BAO observations \cite{Martinelli:2020hud}, while the futuristic case ($N_{\rm lens}=1000$) reaches a sensitivity similar to the one that can be achieved by the combination of the {\it Euclid} BAO survey with the full LSST SNIa survey \cite{Martinelli:2020hud}.

The optimistic case ($N_{\rm lens}=100$) sits somewhere in the middle, but given the reduced number of assumptions made on the cosmological model in the analysis of strongly lensed SNIa, using this approach could allow DDR violation to be disentangled from other cosmological mechanisms \cite{Hogg:2020ktc}.

\begin{table}[!htbp]
\begin{center}
\begin{tabular}{c|c|c} 
\hline
 $N_{\rm lens}=20$ & $N_{\rm lens}=100$ & $N_{\rm lens}=1000$ \\
\hline
\multicolumn{3}{c}{Fiducial $\epsilon_0=0.0$}\\
\hline
$0.0098\pm 0.057$ & $0.015^{+0.019}_{-0.023}$ & $0.0038\pm 0.0065$ \\
\hline
\multicolumn{3}{c}{Fiducial $\epsilon_0=0.01$}\\
\hline
$0.022\pm 0.056$ & $0.025\pm 0.021$ & $0.0127\pm 0.0064$ \\
\hline
\multicolumn{3}{c}{Fiducial $\epsilon_0=0.05$}\\
\hline
$0.056\pm 0.057$ & $0.066\pm 0.022$ & $0.0534\pm 0.0065$ \\
\hline
\hline 
\end{tabular}
\caption{Mean values and $68\%$ confidence level intervals for the $\epsilon_0$ parameter, using mock data with different number of lenses and fiducial values for $\epsilon_0$.}\label{tab:parres}
\end{center}
\end{table}

\subsection{Genetic algorithms \label{subsec:ga}}
Here we describe a non-parametric reconstruction of the duality parameter $\eta(z)$, which is based on a machine learning approach called the Genetic Algorithms (GA) and is complementary to the parameterised analysis of the previous section. The GA are a particular stochastic optimisation approach, loosely inspired from the theory of evolution and mimicking the stochastic operations of mutation, \ie the merging of different individuals to form descendants, and crossover, a random change in the chromosomes of an individual. This is achieved by emulating natural selection, \ie in a given environment, a population (in our case a set of test functions) will evolve and adapt under the pressure of the  operators of mutation and crossover. 

In general, the reproductive success of every member of the population is assumed to be proportional to their fitness, which is a measure of how well they fit the data in question. Here we implement a standard $\chi^2$ statistic as described in the previous sections. For more details on the GA and their applications to cosmology see Refs.~ \cite{Bogdanos:2009ib,Nesseris:2012tt,Nesseris:2010ep,Nesseris:2013bia,Sapone:2014nna, Arjona:2019fwb,Arjona:2020kco,Arjona:2020doi}.

A quick overview of the fitting process is as follows. During the initialisation of the code a set of test functions is formed using a group of orthogonal polynomials, called the grammar. This is a crucial step as it has been shown that the choice of the grammar may significantly affect the convergence rate of the GA code \cite{Bogdanos:2009ib}. Using then this initial population, we encode the duality parameter $\eta(z)$ in every member of the population and we also require that $\eta(z)$ satisfies a set of physical priors and initial conditions. In our analysis we remain completely agnostic regarding the DDR deviation mechanism, so we only assume that the duality parameter satisfies $\eta(z=0)=1$, but we make no assumption of a dark energy model.

After preparing the initial population, we then estimate the fitness of every member using the $\chi^2$ and then we apply the stochastic operators of crossover and mutation to a subset of the best-fitting functions chosen via tournament selection \cite{Bogdanos:2009ib}. We then repeat this process thousands of times, so as to make certain the GA code has converged, and we also use several different random seeds, in order to avoid biasing the run due to a specific random seed.

The errors in the reconstruction are calculated using the path integral approach of Refs.~\cite{Nesseris:2012tt,Nesseris:2013bia}. In this approach the error regions are estimated by integrating the likelihood over all functions of the functional space scanned by the GA. This method has been validated by comparing its error estimates against bootstrap Monte Carlo and Fisher matrix errors \cite{Nesseris:2012tt}. Finally, here we use the publicly available code \texttt{Genetic Algorithms}\footnote{\url{https://github.com/snesseris/Genetic-Algorithms}}.

\begin{figure*}
    \centering
    \begin{tabular}{cc}
    \includegraphics[width=0.49\textwidth]{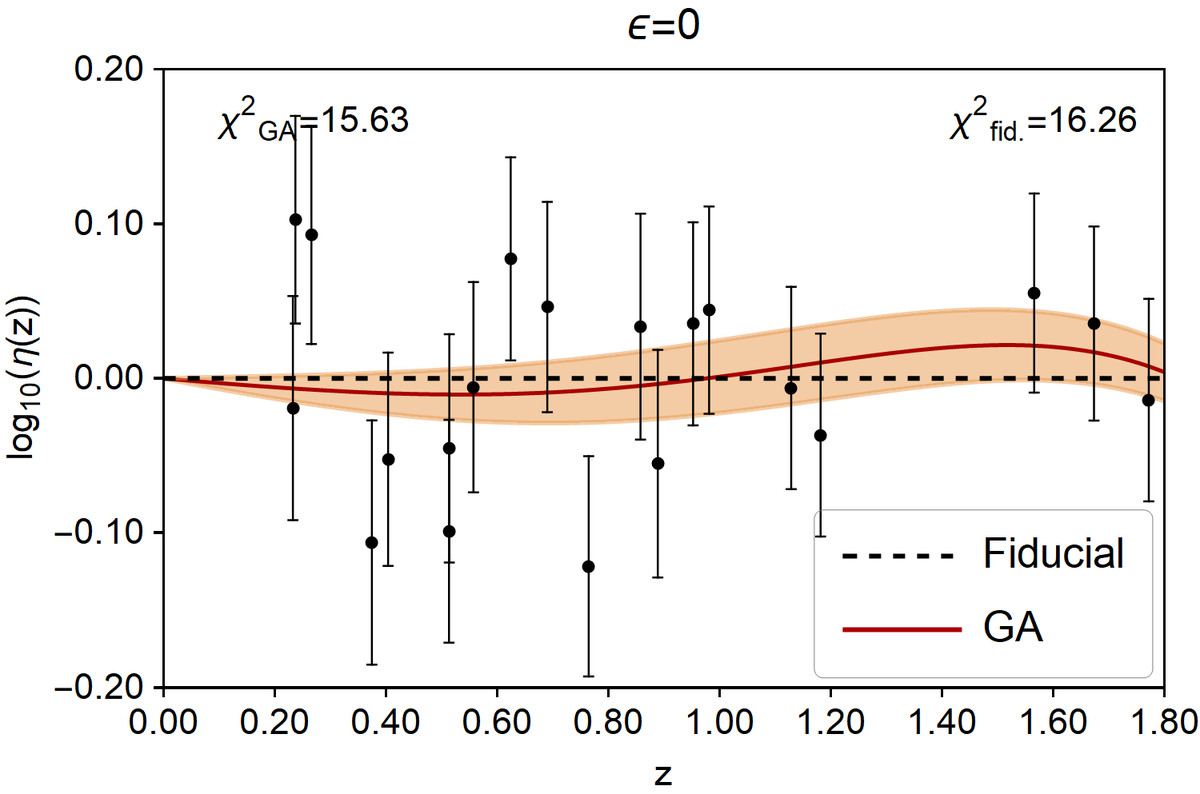} &
    \includegraphics[width=0.49\textwidth]{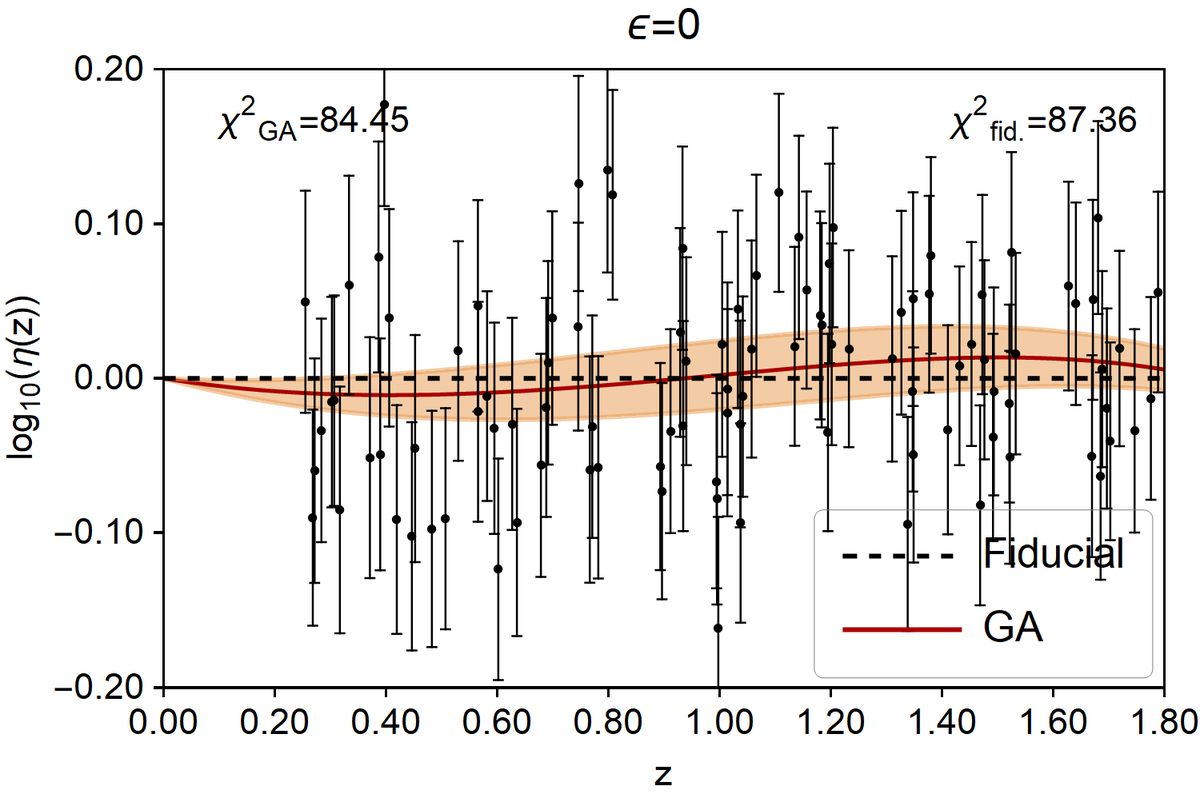}\\
    \includegraphics[width=0.49\textwidth]{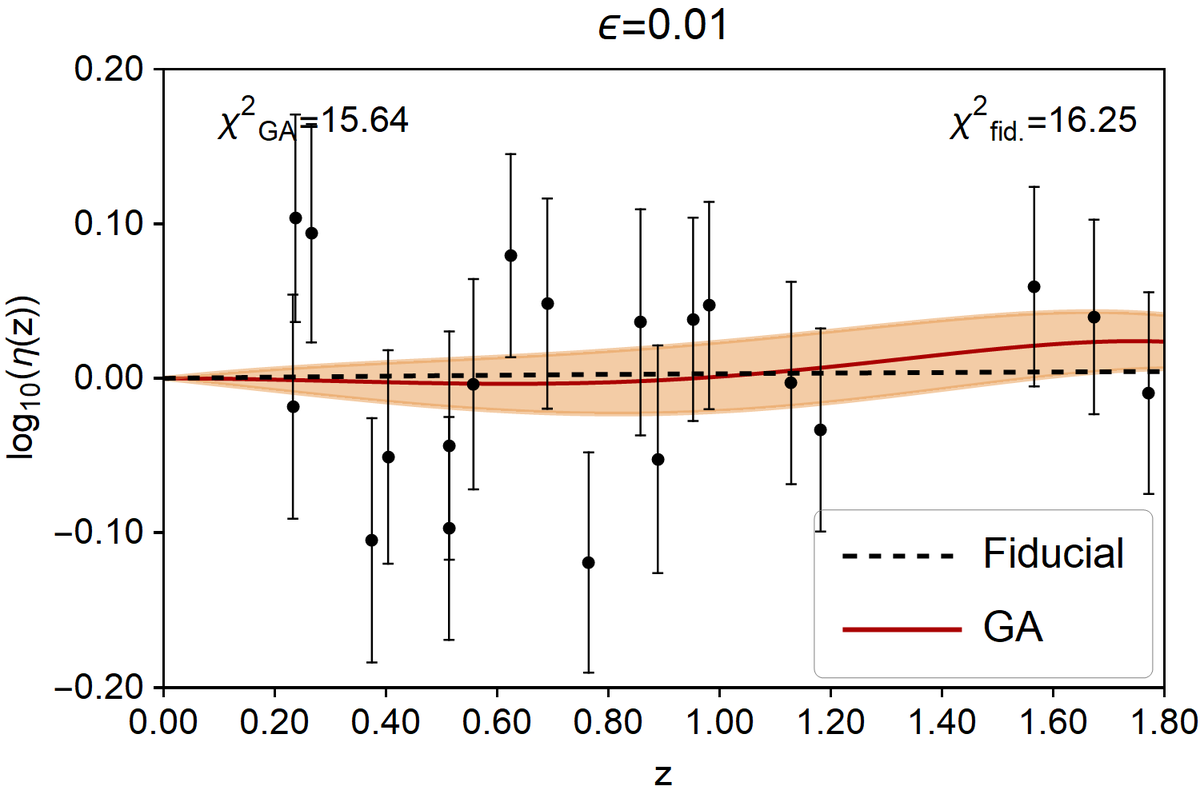} &
    \includegraphics[width=0.49\textwidth]{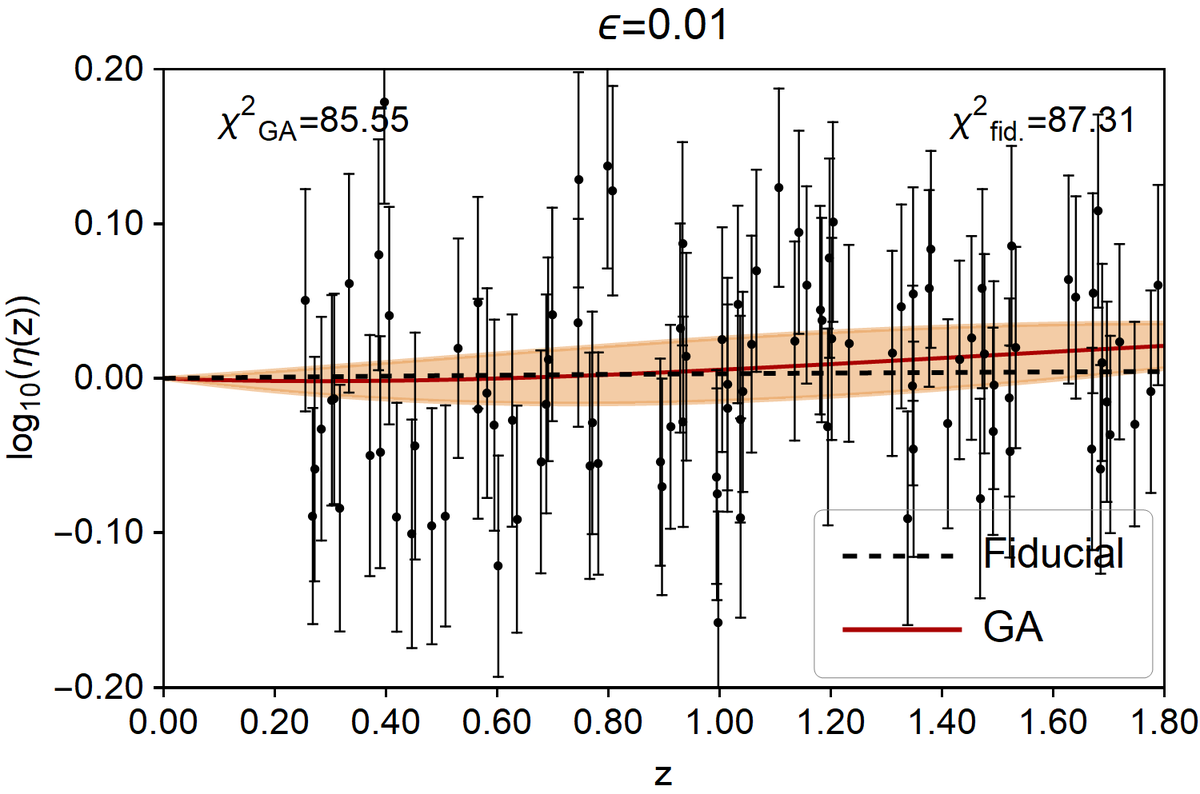}\\
    \includegraphics[width=0.49\textwidth]{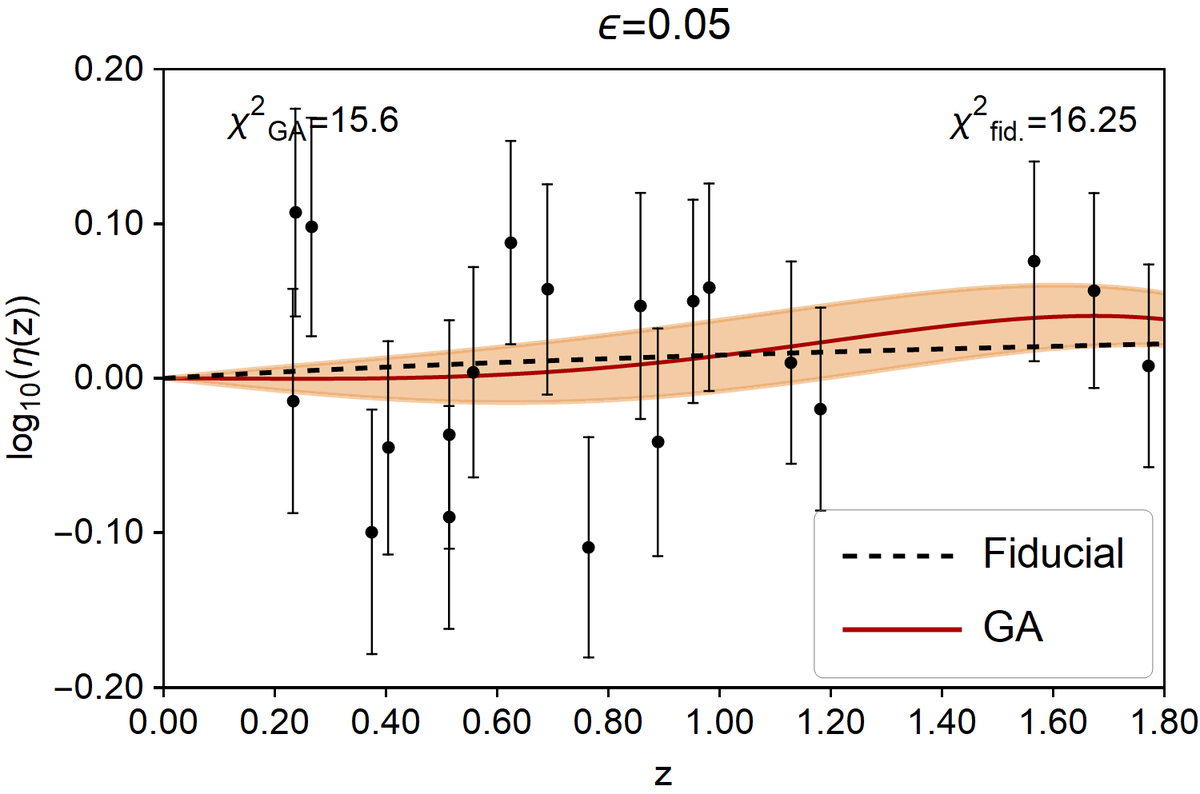} &
    \includegraphics[width=0.49\textwidth]{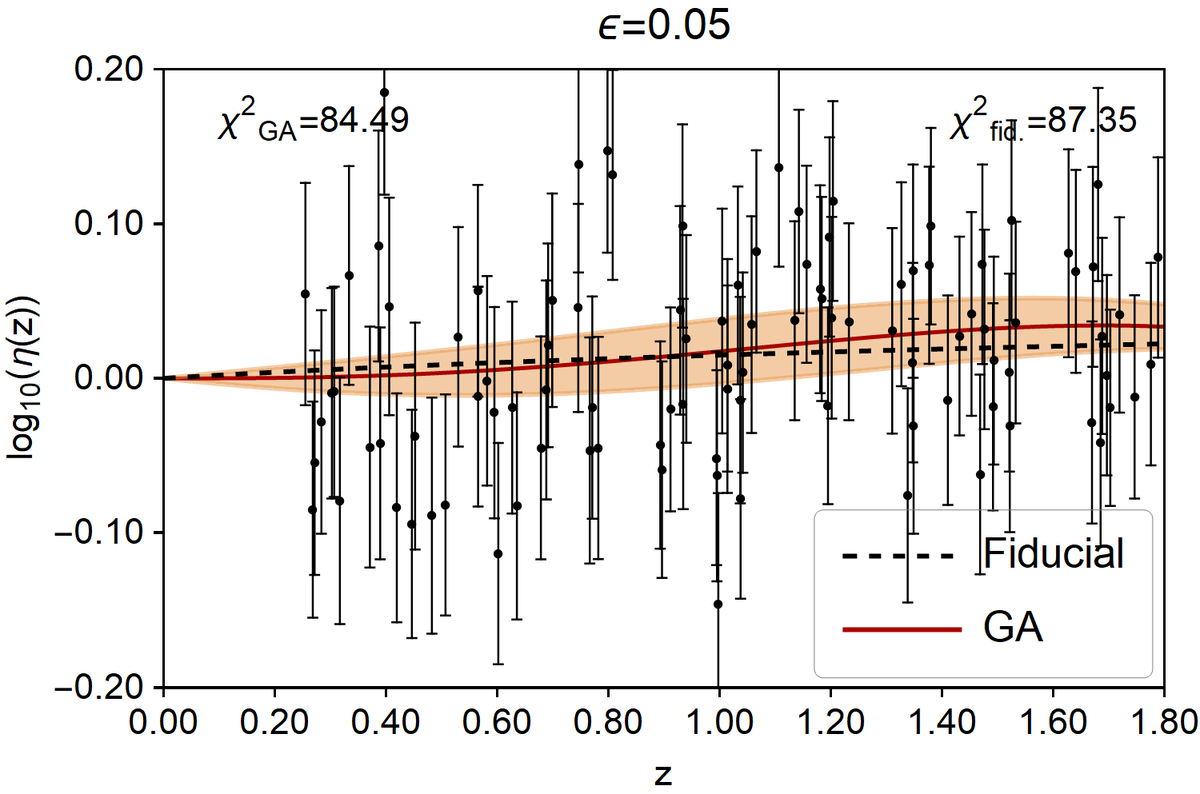}\\
    \end{tabular}
    \caption{The GA reconstructions for the 20 lenses (left column) and for 100 lenses (right column). The mocks in the top row were created with $\epsilon=0$, the ones in the middle row with $\epsilon=0.01$, while the ones in the bottom row with $\epsilon=0.05$. The orange shaded regions show the $1\sigma$ error for the GA, while the dashed black lines show the fiducial model, $\log_{10}(1+z)^{\epsilon_0}$, in each case.}
    \label{fig:garesults}
\end{figure*}

The results of the GA reconstruction can be seen in \autoref{fig:garesults}. In the left column we show the reconstructions for 20 lenses, while in the right column we show the case for 100 lenses. The mocks in the top row were made with $\epsilon=0$, the ones in the middle row with $\epsilon=0.01$, while the ones in the bottom row with $\epsilon=0.05$. As can be seen, in both cases of the 20 and 100 lenses, the GA is able to correctly recover within the errors the underlying fiducial model $\eta_\textrm{fid}(z)=\log_{10}(1+z)^{\epsilon_0}$, shown with a dashed line in each of the panels. 

Specifically, we find that in the case of the 20 lenses the GA is able to predict the fiducial model very well across all redshifts, albeit with a small tension at high redshifts ($z\gtrsim 1.4$) due to the lack of points. On the other hand, in the case of the 100 lenses the GA reconstruction remains very close to the fiducial model at all redshifts.

\subsection{Gaussian processes} \label{subsec:gp}
The classic definition of a Gaussian process (GP) is ``a collection of random variables, any finite number of which have a joint Gaussian distribution'' \cite{Rasmussen}. A GP can be thought of as a generalisation of a Gaussian probability distribution, but whereas a probability distribution describes finite-dimensional random variables, a stochastic process governs the properties of functions. In our case, this function that we use a GP to reconstruct is $\log_{10}{\eta(z)}$, with the redshifts being the input fed to the GP. In general, the GP is completely specified by its mean and covariance functions, though the mean function is usually taken to be zero for the sake of simplicity and a baseline value of zero is hard-coded into many of the popular GP regression packages.

There are many options for the covariance function, or kernel, $k(z,\tilde{z})$. GPs have been applied to reconstruct a wide variety of functions in cosmology (see e.g. \cite{Holsclaw2010a, Holsclaw2010b, Holsclaw2011, Shafieloo:2012ht, Seikel2012, Zhang2018,Martinelli:2019dau, Gerardi:2019obr, Hogg:2020rdp}) and there is still some debate over the best choice of kernel, as the choice can strongly influence the resulting GP reconstruction. In this work, we choose to proceed by tailoring the kernel to one supporting a reconstruction that finds an increasing trend in redshift, as this is what we expect the fiducial models to produce. 

It was found in \cite{Seikel:2013fda} that the Mat\'{e}rn class of kernels performed best when reconstructing the equation of state of dark energy, $w(z)$, using SNIa data. This class of kernels take the following form \cite{Rasmussen}:
\begin{align}
k(z, \tilde{z}) &= \sigma_M^2\frac{2^{1-\nu}}{\Gamma(\nu)} \left(\frac{\sqrt{2 \nu}d(z,\tilde{z})}{\ell} \right)^{\nu} \nonumber  \\ 
 &\times \ K_\nu \left( \frac{\sqrt{2 \nu}d(z,\tilde{z})}{\ell}\right) \label{eq:matern},
\end{align}
where $d(z,\tilde{z})$ represents the Euclidean distance between the inputs $z$ and $\tilde{z}$, $\Gamma(\nu)$ is the gamma function, $K_\nu$ is a modified Bessel function and $\nu$ controls the shape of the covariance function, tending to the Gaussian limit as $\nu \rightarrow \infty$. The hyperparameters $\ell$ and $\sigma_M$ correspond to the approximate length scale over which the function varies and the magnitude of those variations respectively. The choice of a half-integer value for $\nu$ is made in order to remove the dependence on the Bessel function \cite{Seikel:2013fda}. The larger the value of $\nu$, the smoother the resulting GP, although for $\nu \geq 7/2$, the results become hard to distinguish from one another \cite{Rasmussen}. Overall, this makes $\nu = 5/2$ a good choice.

In the course of our analysis, we found that when a Mat\'{e}rn kernel is used alone, the GP struggles to follow the trend in redshift introduced by the fiducial models of $\epsilon=0.01$ and $\epsilon=0.05$. We therefore create a custom kernel that better suits our problem, by adding a dot product kernel to a Mat\'{e}rn ($\nu =5/2$) kernel. The dot product kernel takes the general form
\begin{equation}
    k(z, \tilde{z}) = \sigma_d + z \cdot \tilde{z}, \label{eq:dotproduct}
\end{equation}
where the hyperparameter $\sigma_d$ acts on the dot product kernel in a similar way to how $\sigma_M$ acts on the Mat\'{e}rn kernel. For the Mat\'{e}rn class of kernels, $\sigma_M$ acts to rescale the GP covariance, whereas for the dot product kernel, $\sigma_d$ acts as a constant offset of the covariance of the GP. We note that the dot product kernel is non-stationary, meaning that the resulting GP depends not only on the relative positions of the points, but on their absolute positions. A translation in the input space (i.e. shifting the mock data points in redshift) will therefore result in a different GP prediction from the dot product kernel even if the kernel hyperparameter is kept fixed \cite{Duvenaud2014}.

We use the Gaussian process regressor provided by the Python package \texttt{scikit-learn} \cite{scikit-learn} to perform our reconstruction of $\log_{10}{\eta(z)}$ with the custom kernel described above. The package also allows for optimisation of the value of any hyperparameters in the kernel by maximising the log-likelihood of the GP output. We list the optimised values of $\ell$, $\sigma_M$ and $\sigma_d$ in \autoref{tab:hyperparam} to give an idea of the general behaviour of our custom kernel. 

Note that we do not fix these values by hand in the kernel. The only information we give to the kernel is the upper and lower bound that the optimiser explores between for the value of the length scale $\ell$. This choice of bound can have an effect on the resulting reconstruction, as there may be multiple values of the hyperparameters that maximise the log-likelihood. However, the optimisation routine will only be able to find one of the maximal values each time the procedure is run. The bounds can therefore be manually shrunk to eliminate all but one of the maximal values of each of the hyperparameters, forcing the GP to use that particular combination. 

The value of the hyperparameter $\ell$ corresponds to the average variation in the $z$-direction of the data, and is expected to be of order of the average distance between each mock data point. Therefore,to select the upper and lower bounds for the length scale in the Mat\'{e}rn kernel, we considered the approximate average distance between each mock data point in the catalogue, roughly $0.08$ in terms of the redshift in the case of 20 lenses. Since it is squared, we then expect the learned length scale to be of the order $10^{-3}$. In the case of 100 lenses, the mock data points are spaced closer together, leading us to expect a learned length scale on the order of $10^{-4}$. We therefore set the bounds of the Mat\'{e}rn kernel as $10^{-5}$ and $10^{-1}$ to safely incorporate these expected values.

The value of $\sigma_M$ instead corresponds to the typical variation in amplitude of the function, which is expected to be of the order of the average error of the data points \emph{i.e.} $\sim 0.05$. Finally, the dot product kernel is equivalent to a linear regression in which $\sigma_d$ is the intercept of the fit. From \autoref{eq.etath} it is straightforward to see that $\sigma_d \approx \epsilon^2_0 = O(10^{-4})$. We therefore see that the expected values for $\sigma_d$ and $\sigma_M$ fall well within the imposed bounds for the GP hyperparameters. While at first glance this ``recipe'' used to build the kernel appears somewhat na\"{i}ve, its validity is confirmed by the optimised hyperparameter values reported in \autoref{tab:hyperparam}. 

\begin{figure*}
    \centering
    \begin{tabular}{cc}
    \includegraphics[width=0.485\textwidth]{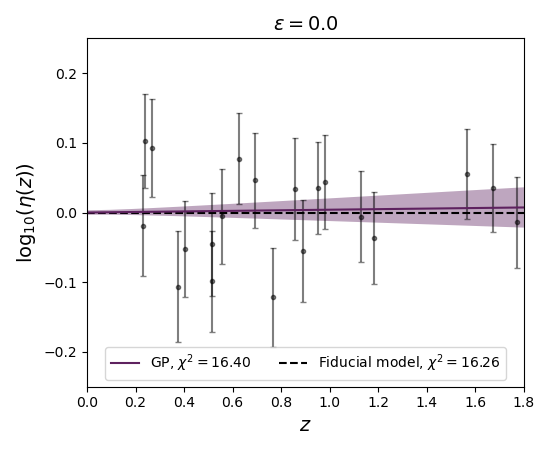} &
    \includegraphics[width=0.485\textwidth]{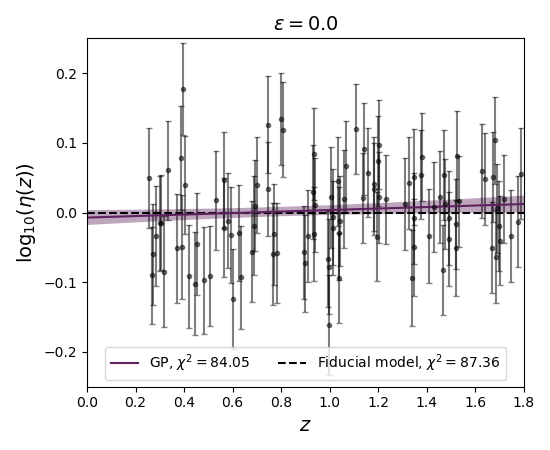}\\
    \includegraphics[width=0.485\textwidth]{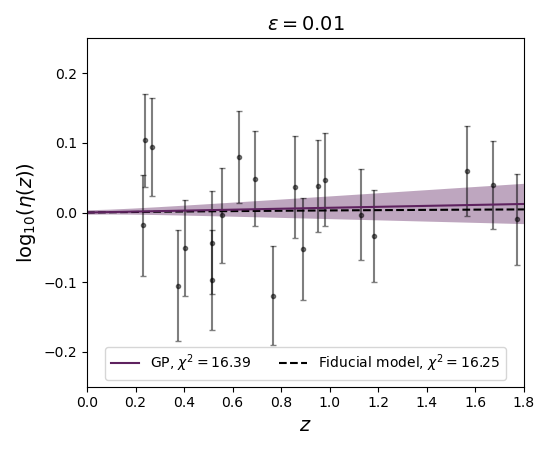} &
    \includegraphics[width=0.485\textwidth]{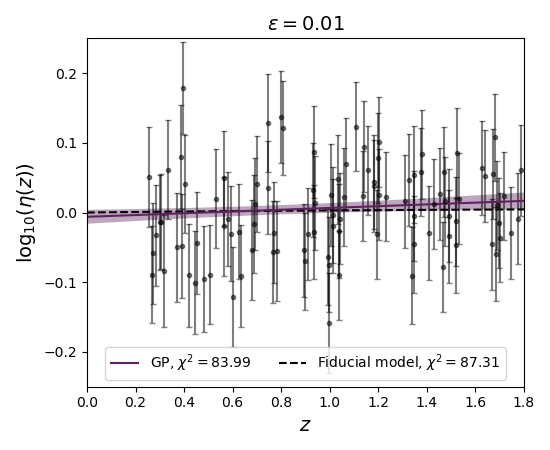}\\
    \includegraphics[width=0.485\textwidth]{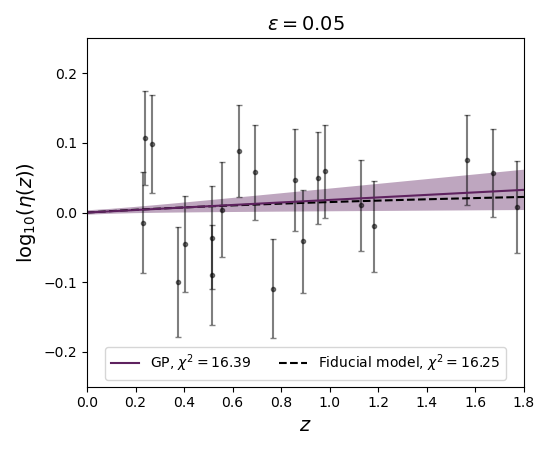} &
    \includegraphics[width=0.485\textwidth]{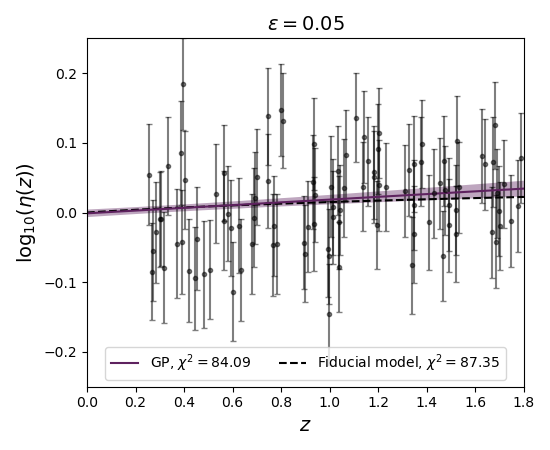}\\
    \end{tabular}
    \caption{The GP reconstructions for the 20 lenses (left column) and for 100 lenses (right column). The mocks in the top row were created with $\epsilon_0=0$, the ones in the middle row with $\epsilon_0=0.01$, while the ones in the bottom row with $\epsilon_0=0.05$. The shaded regions show the $1\sigma$ error for the GP, while the dashed black lines show the fiducial model, $\log_{10}(1+z)^{\epsilon_0}$, in each case. }
    \label{fig:gpresults}
\end{figure*}

The results of the GP reconstruction using the custom kernel are shown in \autoref{fig:gpresults}. The left column shows the reconstructions of $\log_{10}{\eta(z)}$ for the realistic case of 20 strongly lensed SNIa, and the right column shows the optimistic case of 100 lenses. The mock data in the top row was created with no deviation from $\Lambda$CDM or the standard DDR, i.e. $\epsilon=0.0$, while the middle row shows the mock data for which $\epsilon=0.01$ and the bottom row $\epsilon=0.05$. 

In the realistic case of 20 lenses, we see that the relatively small number of points does not prevent the GP from correctly recovering the fiducial model (dashed line in all three panels of \autoref{fig:gpresults}) to within $1\sigma$ for all the fiducial cases. 

In the optimistic case of 100 lenses, the error of the GP at high redshift is decreased with respect to the 20 lens case, due to the increased information given to the GP by the additional mock data points. However, for this particular mock dataset realisation, the reconstruction does not recover the fiducial model as well as the 20 lens case, with a slight overestimation of the $\log_{10}\eta(z)$ function at higher redshifts for all three values of $\epsilon_0$. However, even with this overestimation, the reconstruction is again never more than $1\sigma$ away from the true fiducial model.

In all cases we report the $\chi^2$ statistic for the fiducial model and the GP reconstruction in the legend of the plots.

\begin{table}[!htbp]
\begin{center}
\begin{tabular}{Sl|l|l|l} % S prefix allows cells to be stretched by cellspace
\hline
\multicolumn{1}{c|}{$\epsilon_0$} & \multicolumn{1}{|c|}{$\ell$} & \multicolumn{1}{|c|}{$\sigma_M$} & \multicolumn{1}{|c}{$\sigma_d$} \\
\hline
\multicolumn{4}{c}{$N_{\rm lens}=20$}\\  
\hline
0.0 &  $1.00 \times 10^{-3}$  & $3.16 \times 10^{-3}$ & $1.47 \times 10^{-6}$ \\
0.01 & $1.00 \times 10^{-3}$  & $3.16 \times 10^{-3}$ & $1.12\times 10^{-6}$ \\
0.05 & $1.00 \times 10^{-3}$  & $3.16 \times 10^{-3}$ & $1.04 \times 10^{-6}$ \\
\hline
\multicolumn{4}{c}{$N_{\rm lens}=100$}\\ 
\hline
% $\epsilon$ & $\ell$ & $\sigma_M$ & $\sigma_d$ \\
% \hline
0.0  & $1.00 \times 10^{-3}$ & $3.16 \times 10^{-3}$ & $1.25\times 10^{-2}$ \\
0.01 & $1.00 \times 10^{-3}$ & $3.16 \times 10^{-3}$ & $1.13\times 10^{-2}$ \\
0.05 & $1.00 \times 10^{-3}$ & $3.16 \times 10^{-3}$ & $4.56 \times 10^{-3}$ \\
\hline
\hline 
\end{tabular}
\caption{Values of the kernel hyperparameters after optimisation.  
}\label{tab:hyperparam}
\end{center}
\end{table}

\section{Conclusions\label{sec:conclusions}} 
In this paper we investigated the possibility of using future observations of strongly lensed Type Ia supernovae to constrain deviations from the standard distance duality relation. A departure from the DDR could be a significant smoking gun for deviations from the standard cosmological model, as it would signal that fundamental assumptions are violated, which we discussed in \autoref{sec:theory}.

Such violations are usually investigated in the literature by combining different observations together; this allows the luminosity and angular distances to be reconstructed separately and the function $\eta(z)$, equal to unity in the standard model, to be constrained. In \autoref{sec:lensedsnia} we discussed how the observation of strongly lensed SNIa can instead directly provide the two distances at the redshift of the source, and can therefore be used to obtain measurements of $\eta(z)$, avoiding the need to reconstruct the two distances. Notice however that such a measurement is possible only under certain assumptions; one needs to be able to obtain the luminosity distance of the lensed supernovae and remove any possible magnification due to the lens, while the measurement of the angular distance at the source redshift can be obtained from the time delay distance only through the assumption of a flat Universe and if kinematic measurements of the lens galaxy are available.

Other than these assumptions, the use of such observations allows us to obtain our results without any further dependence on the cosmological model, even in the parametric approach that we discuss in \autoref{sec:analysis}. For this case we find that, as expected, the results strongly depend on the number of systems that will be observed by future surveys; for a realistic number of strongly lensed SNIa ($N_{\rm lens}=20$) the constraints we obtain on $\epsilon_0$ are of the order of those obtained through the combination of currently available SNIa and BAO surveys, while in our most futuristic case ($N_{\rm lens}=1000$) bounds on DDR violation obtained through strong lensing are expected to be competitive with those forecast for upcoming LSS surveys.

The results of the Genetic Algorithm reconstruction for both cases of 20 and 100 lenses for $\epsilon_0=(0.0, 0.01, 0.05)$ were shown in \autoref{subsec:ga} and in \autoref{fig:garesults}. In all cases the GA was able to correctly recover the underlying fiducial model within the errors.

In \autoref{subsec:gp}, we presented the results of our Gaussian process reconstruction. We reconstructed $\log_{10}\eta(z)$ for the fiducial models of $\epsilon_0=0.0$, $\epsilon_0=0.01$ and $\epsilon_0=0.05$ using both 20 lenses and 100 lenses, finding that the GP was well able to correctly recover the underlying fiducial in the mock data.

In summary, we have shown how strongly lensed SNIa will be a powerful probe of distance measures in cosmology in the upcoming LSST era. We have discussed how these systems are uniquely able to provide measurements of both luminosity and angular diameter distances, allowing excellent constraints to be placed on the distance duality relation. If any deviations from this relation were to be detected it would be an exciting hint at possible new physics easily accessible to other next-generation surveys.

%I need vacations...

\section*{CRediT authorship contribution statement}
\textbf{Fabrizio Renzi:} Methodology, Software, Formal analysis, Validation, Writing -- original draft. \textbf{Natalie B. Hogg:} Software, Formal analysis, Writing -- original draft, Writing - Review \& Editing. \textbf{Matteo Martinelli:} Conceptualization, Software, Formal analysis, Validation, Writing -- original draft, Supervision. \textbf{Savvas Nesseris:} Conceptualization, Software, Formal analysis, Writing -- original draft.

\section*{Declaration of competing interest}

The authors declare that they have no known competing financial interests or personal relationships that could have appeared to influence the work reported in this paper.

\section*{Acknowledgements}
We thank Mike Shengbo Wang for a useful discussion regarding the statistical validity of the mock data. Numerical computations were done on the Hydra HPC Cluster of the Instituto de F\'isica Te\'orica UAM/CSIC. 
f
FR acknowledges support from the NWO and the Dutch Ministry of Education, Culture and Science (OCW), and from the D-ITP consortium, a program of the NWO that is funded by the OCW. NBH is supported by UK STFC studentship ST/N504245/1. MM has received the support of a fellowship from ``la Caixa” Foundation (ID 100010434), with fellowship code LCF/BQ/PI19/11690015, and the support of the Spanish Agencia Estatal de Investigacion through the grant “IFT Centro de Excelencia Severo Ochoa SEV-2016-0597”. MM also wants to thank the Big Star Bar for providing a work space and an internet connection during this period of remote work. SN acknowledges support from the research projects PGC2018-094773-B-C32, the Centro de Excelencia Severo Ochoa Program SEV-2016-059 and the Ram\'{o}n y Cajal program through Grant No. RYC-2014-15843. 

In this work we made use of the following Python packages that are not mentioned in the text: GetDist \cite{GetDist}, a tool for the analysis of MCMC samples, Matplotlib \cite{Matplotlib}, for the realisation of the plots in the paper, NumPy \cite{NumPy}, for numerical linear algebra, and SciPy \cite{2020SciPy-NMeth}, for numerical sampling of the statistical distributions involved in our data analysis.

\appendix
\section{Details of the mock catalogue creation} \label{sec:appendixA}
\begin{figure}
    \centering
    \includegraphics[width=0.49\textwidth]{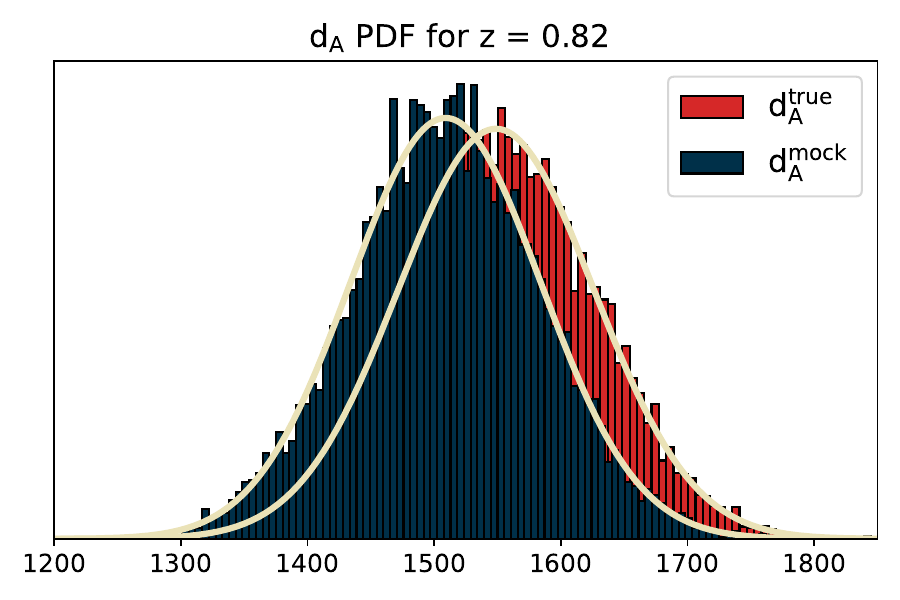}\\
    \caption{Comparison of the PDFs obtained from the truth value of the angular diameter distance $d_A^{\rm true}$ and the corresponding distribution for the mock value $d_A^{\rm mock}$ at fixed redshift. The solid lines show the corresponding theoretical Gaussian PDFs with $\sigma_{D_i} = 0.05$}
    \label{fig:da_pdf}
\end{figure}

In this Appendix we describe in more detail the MCMC-like approach used to construct our mock catalogues of $\eta(z_i)$ with $i = 1 \dots N_{\rm lens}$. As discussed in the main text, the methodology followed to generate our mock catalogues has three distinct steps. We start by constructing the probability distribution function (PDF) of the distances involved in the DDR. For a given redshift $z_i$ we start drawing random Gaussian deviates, $ \delta D(z_i)$, from a Gaussian distribution of the form:
%we extract the value of the mean of PDF of $D_{i, \rm mock}$ from a Gaussian distribution,
%\begin{equation}
%\bar{D}_i = \mathcal{N}(D^{\rm true}_i,\sigma_{D_i}D^{\rm true}_i)\, ,
%\end{equation}

\begin{equation}
    \mathcal{N}(0,\sigma_{D_i}D^{\rm true}_i)
\end{equation}
with $D^{\rm true}_i$ being the true value of the distance for an assumed cosmological model (for this work, $H_0 = 70$ km s$^{-1}$ Mpc$^{-1}$ and $\Omega_m = 0.3$  along with the three chosen values of $\epsilon_0$) at $z_i$ and $\sigma_{D_i}$, the observational error on this distance.
We then construct the PDF of $D_{i,\rm mock}$ by extracting 10,000 samples from a Gaussian distribution with mean $\bar{D}_i = D^{\rm true}_i + \delta D_i$ and standard deviation $\sigma_{D_i}$ \ie 
\begin{equation}
D_{i,\rm mock} = \mathcal{N}(\bar{D}_{i},\sigma_{D_i}\bar{D}_{i})\, .
\end{equation}
A comparison of the true and mock PDFs is plotted in \autoref{fig:da_pdf} for the angular diameter distance. 

With the PDFs of $d_A$, $d_{\Delta}$ and $\mu$ in hand, we proceed in an MCMC-like fashion. We assume the PDFs of $d_A$, $d_{\Delta t}$ and $\mu$ to be the posteriors of a hypothetical MCMC run with the three distances as independent parameters, so that at each redshift $z_i$, each triplet $\{d_{A,n}, d_{\Delta t, n}, \mu_{n}\ |\ n = 1 \dots 10^4 \}$ constitutes a sample of an MCMC chain. Therefore at each $n$ we combine the triplet values, using \autoref{eq:etaTD} to obtain a sample of the posterior of $(\log_{10}\eta(z_i))_n$, \ie we treat $\log_{10}\eta(z_i)$ as a derived parameter of the MCMC. We apply this procedure to all 10,000 samples to construct the distribution of $\log_{10}\eta(z_i)$. 

\begin{figure}
    \centering
    \includegraphics[width=0.49\textwidth]{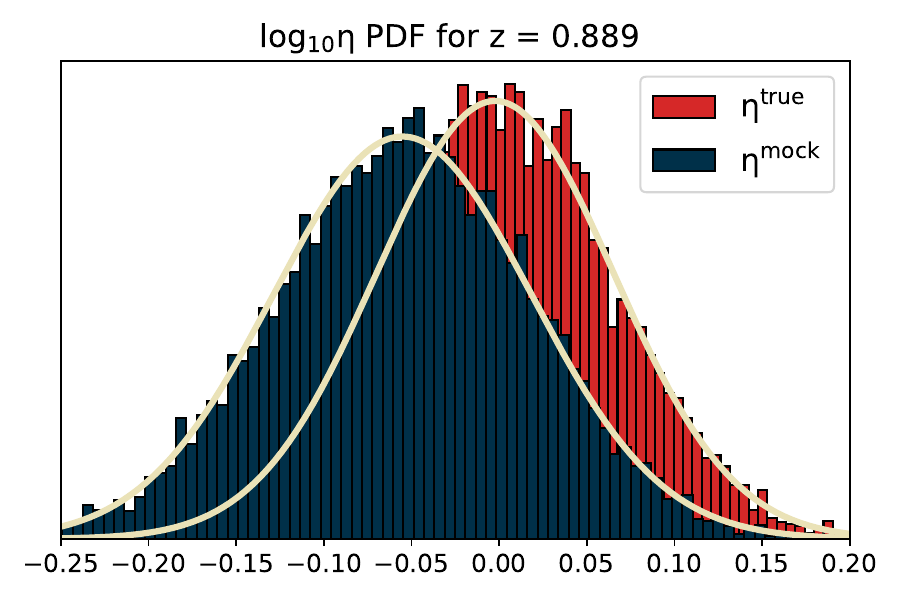}\\
    \caption{Comparison of the PDFs obtained from the truth value of the DDR function $\log_{10}\eta^{\rm true}$ and the corresponding distribution for the mock value $\log_{10}\eta^{\rm mock}$ at fixed redshift. The solid lines show the corresponding theoretical Gaussian PDFs with $(\sigma_{\log_{10}\eta})_{\rm mock} = 0.068$ and $(\sigma_{\log_{10}\eta})_{\rm true} = 0.074$.}
    \label{fig:eta_pdf}
\end{figure}

\begin{figure}
    \centering
    \includegraphics[width=0.49\textwidth]{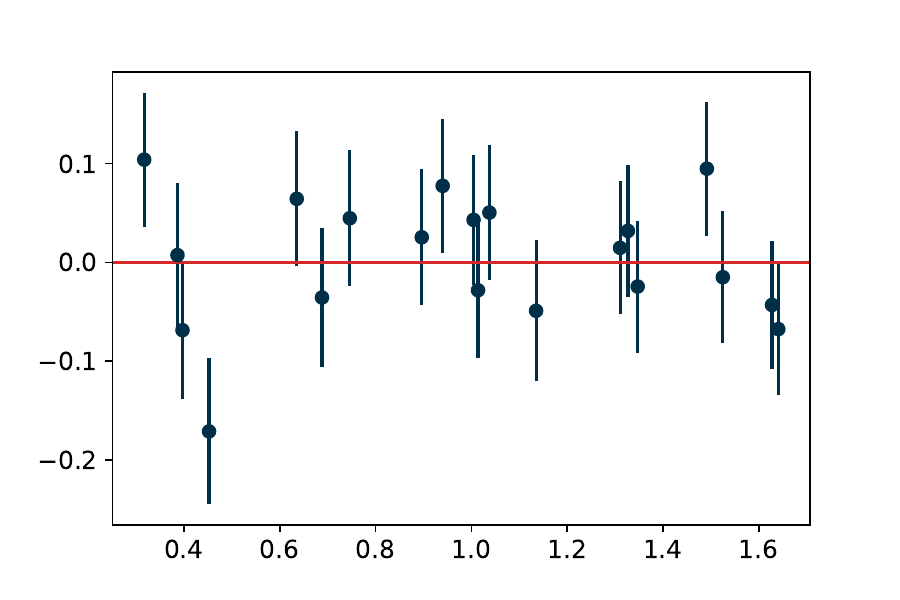}\\
    \caption{A sample mock for the 20 lenses catalogue constructed with the methodology described in \ref{sec:appendixA}.}
    \label{fig:eta_mock}
\end{figure}

A comparison of the true and mock PDFs of $\log_{10}\eta(z_i)$ is plotted in \autoref{fig:eta_pdf} while in \autoref{fig:eta_mock} we show a sample mock for $N_{\rm lenses} = 20$. As we can see from \autoref{fig:eta_pdf}, the assumption $\log_{10}\eta(z_i) \approx \mathcal{N}(0,\sigma_{\log_{10}\eta(z_i)})$ is very much in agreement with the numerical distributions of $\log_{10}\eta(z_i)$ constructed with our methodology.

\begin{figure}
    \centering
    \includegraphics[width=0.49\textwidth]{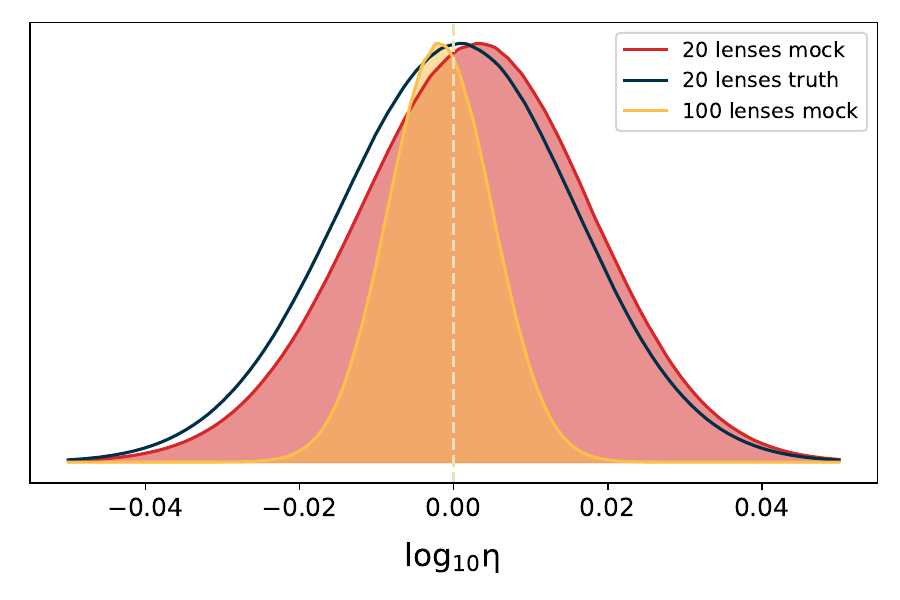}\\
    \caption{The combined PDFs of $\log_{10}\eta$ for mocks of $N_{\rm lens} = 20,\, 100$ plotted against the combined true PDFs of $\log_{10}\eta$ for $N_{\rm lens} = 20$.  }
    \label{fig:eta_post_comparison}
\end{figure}

\begin{figure}
    \centering
    \includegraphics[width=0.49\textwidth]{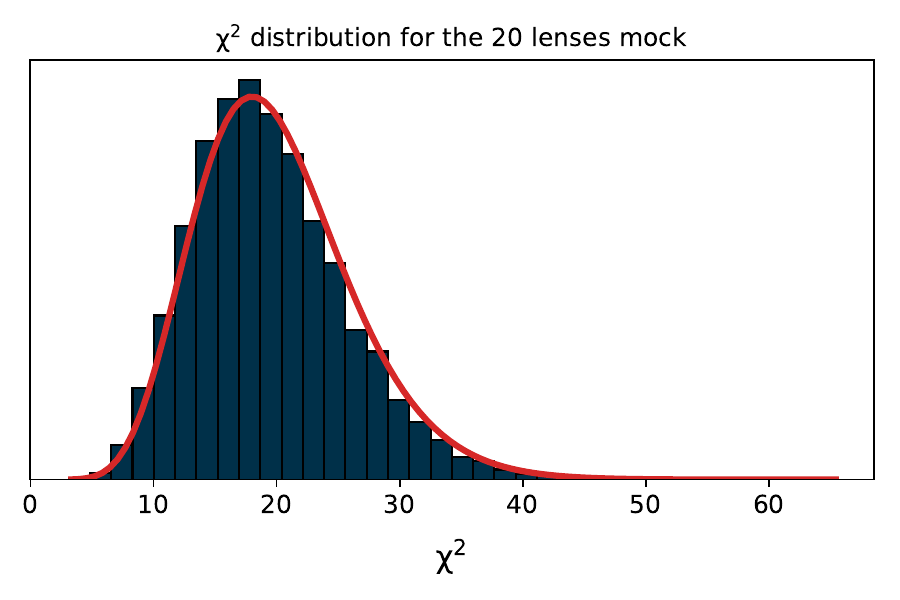}\\
    \caption{Comparison between the distribution of $\chi^2$ values for the 20 lens mock dataset (dark histogram) with the theoretical $\chi^2$ distribution for 20 degrees of freedom (red solid line).}
    \label{fig:chi2}
\end{figure}

From the PDFs of $\log_{10}\eta(z)$, we can also perform some sanity checks. First of all, assuming that $\log_{10}\eta(z) = const $, we can multiply the PDFs of all the $\log_{10}\eta(z_i)$ to obtain a combined posterior and therefore the mock best fit for $\log_{10}\eta(z)$. We show the combined PDFs of $\log_{10}\eta$ for two mocks of $N_{\rm lens} = 20,\, 100$ plotted against the combined true PDFs of $\log_{10}\eta$ for $N_{\rm lens} = 20$ in \autoref{fig:eta_post_comparison}.
While this best-fit value will not be as accurate as the one obtained from a full MCMC sampling, it can signal inconsistency in the mock dataset without the need for a complex analysis.
Furthermore, we can construct the $\chi^2$ distribution, testing 10,000 realisations of a mock against the hypothesis $\log_{10}\eta(z_i) \approx \mathcal{N}(0,\sigma_{\log_{10}\eta(z_i)})$ as an additional sanity check. In \autoref{fig:chi2} we show the comparison between the distribution of $\chi^2$ values for the 20 lens mock dataset and the theoretical $\chi^2$ distribution for 20 degrees of freedom. We can see that the mock distribution follows the theoretical one extremely well.

So far, we found that our mocks are generally within the $1\sigma$ bounds of the true combined PDF, even though a significant deviation from the fiducial might happen in correspondence with the higher/lower tail of the $\chi^2$ distribution for the mocks.
In summary, this procedure has two main advantages: (1) it exposes the PDFs of the data points of the mocks, allowing them to be used for sanity checks and eventually for a full MCMC sampling similar to what has been done for the analysis of the H0LiCOW lenses (see e.g. \cite{Wong:2019kwg}) and (2) it allows us to reconstruct the errors of the data points directly from their posteriors, removing any assumptions coming from the standard error propagation formula.

\bibliographystyle{elsarticle-num-names}
\bibliography{stronglensing}

\begin{thebibliography}{81}
\expandafter\ifx\csname natexlab\endcsname\relax\def\natexlab#1{#1}\fi
\providecommand{\url}[1]{\texttt{#1}}
\providecommand{\href}[2]{#2}
\providecommand{\path}[1]{#1}
\providecommand{\DOIprefix}{doi:}
\providecommand{\ArXivprefix}{arXiv:}
\providecommand{\URLprefix}{URL: }
\providecommand{\Pubmedprefix}{pmid:}
\providecommand{\doi}[1]{\href{http://dx.doi.org/#1}{\path{#1}}}
\providecommand{\Pubmed}[1]{\href{pmid:#1}{\path{#1}}}
\providecommand{\bibinfo}[2]{#2}
\ifx\xfnm\relax \def\xfnm[#1]{\unskip,\space#1}\fi
%Type = Article
\bibitem[{Suyu et~al.(2020)}]{Suyu:2020opl}
\bibinfo{author}{S.~H. Suyu}, et~al.,
\newblock \bibinfo{title}{{HOLISMOKES -- I. Highly Optimised Lensing
  Investigations of Supernovae, Microlensing Objects, and Kinematics of
  Ellipticals and Spirals}},
\newblock \bibinfo{journal}{Astron. Astrophys.} \bibinfo{volume}{644}
  (\bibinfo{year}{2020}) \bibinfo{pages}{A162}.
  \DOIprefix\doi{10.1051/0004-6361/202037757}.
  \href{http://arxiv.org/abs/2002.08378}{{\tt arXiv:2002.08378}}.
%Type = Book
\bibitem[{Schneider et~al.(1992)Schneider, Ehlers, and
  Falco}]{Grav_lensing1992}
\bibinfo{author}{P.~Schneider}, \bibinfo{author}{J.~Ehlers},
  \bibinfo{author}{E.~Falco}, \bibinfo{title}{Gravitational Lenses},
  \bibinfo{publisher}{Springer}, \bibinfo{year}{1992}.
%Type = Book
\bibitem[{Schneider et~al.(2006)Schneider, Kochanek, and
  Wambsganss}]{Grav_lensing2006}
\bibinfo{author}{P.~Schneider}, \bibinfo{author}{C.~S. Kochanek},
  \bibinfo{author}{J.~Wambsganss}, \bibinfo{title}{Gravitational Lensing:
  Strong, Weak and Micro}, \bibinfo{publisher}{Springer}, \bibinfo{year}{2006}.
%Type = Article
\bibitem[{Borra(1997)}]{Borra1997}
\bibinfo{author}{E.~F. Borra},
\newblock \bibinfo{title}{Detection of gravitational lenses and measurement of
  time delays from classical electromagnetic radiation fluctuations},
\newblock \bibinfo{journal}{Monthly Notices of the Royal Astronomical Society}
  \bibinfo{volume}{289} (\bibinfo{year}{1997}) \bibinfo{pages}{660--664}.
  \URLprefix \url{http://arxiv.org/abs/astro-ph/9704074}.
  \href{http://arxiv.org/abs/Arxiv:astro-ph/9704074v1}{{\tt
  arXiv:Arxiv:astro-ph/9704074v1}}.
%Type = Article
\bibitem[{Borra(2008)}]{Borra2008}
\bibinfo{author}{E.~F. Borra},
\newblock \bibinfo{title}{Observations of time delays in gravitational lenses
  from intensity fluctuations: The coherence function},
\newblock \bibinfo{journal}{Monthly Notices of the Royal Astronomical Society}
  \bibinfo{volume}{389} (\bibinfo{year}{2008}) \bibinfo{pages}{364--370}.
  \URLprefix \url{http://arxiv.org/abs/0806.2252}.
  \href{http://arxiv.org/abs/Arxiv:0806.2252v1}{{\tt arXiv:Arxiv:0806.2252v1}}.
%Type = Article
\bibitem[{Treu(2010)}]{Treu:2010uj}
\bibinfo{author}{T.~Treu},
\newblock \bibinfo{title}{{Strong Lensing by Galaxies}},
\newblock \bibinfo{journal}{Ann. Rev. Astron. Astrophys.} \bibinfo{volume}{48}
  (\bibinfo{year}{2010}) \bibinfo{pages}{87--125}.
  \DOIprefix\doi{10.1146/annurev-astro-081309-130924}.
  \href{http://arxiv.org/abs/1003.5567}{{\tt arXiv:1003.5567}}.
%Type = Article
\bibitem[{Treu and Marshall(2016)}]{Treu:2016ljm}
\bibinfo{author}{T.~Treu}, \bibinfo{author}{P.~J. Marshall},
\newblock \bibinfo{title}{{Time Delay Cosmography}},
\newblock \bibinfo{journal}{Astron. Astrophys. Rev.} \bibinfo{volume}{24}
  (\bibinfo{year}{2016}) \bibinfo{pages}{11}.
  \DOIprefix\doi{10.1007/s00159-016-0096-8}.
  \href{http://arxiv.org/abs/1605.05333}{{\tt arXiv:1605.05333}}.
%Type = Article
\bibitem[{Suyu et~al.(2018)Suyu, Chang, Courbin, and Okumura}]{Suyu:2018vqs}
\bibinfo{author}{S.~H. Suyu}, \bibinfo{author}{T.-C. Chang},
  \bibinfo{author}{F.~Courbin}, \bibinfo{author}{T.~Okumura},
\newblock \bibinfo{title}{{Cosmological distance indicators}},
\newblock \bibinfo{journal}{Space Sci. Rev.} \bibinfo{volume}{214}
  (\bibinfo{year}{2018}) \bibinfo{pages}{91}.
  \DOIprefix\doi{10.1007/s11214-018-0524-3}.
  \href{http://arxiv.org/abs/1801.07262}{{\tt arXiv:1801.07262}}.
%Type = Article
\bibitem[{Shiralilou et~al.(2020)Shiralilou, Martinelli, Papadomanolakis,
  Peirone, Renzi, and Silvestri}]{Shiralilou:2019div}
\bibinfo{author}{B.~Shiralilou}, \bibinfo{author}{M.~Martinelli},
  \bibinfo{author}{G.~Papadomanolakis}, \bibinfo{author}{S.~Peirone},
  \bibinfo{author}{F.~Renzi}, \bibinfo{author}{A.~Silvestri},
\newblock \bibinfo{title}{{Strong Lensing Time Delay Constraints on Dark
  Energy: a Forecast}},
\newblock \bibinfo{journal}{JCAP} \bibinfo{volume}{04} (\bibinfo{year}{2020})
  \bibinfo{pages}{057}. \DOIprefix\doi{10.1088/1475-7516/2020/04/057}.
  \href{http://arxiv.org/abs/1910.03566}{{\tt arXiv:1910.03566}}.
%Type = Article
\bibitem[{Birrer et~al.(2020)}]{Birrer:2020tax}
\bibinfo{author}{S.~Birrer}, et~al.,
\newblock \bibinfo{title}{{TDCOSMO - IV. Hierarchical time-delay cosmography
  \textendash{} joint inference of the Hubble constant and galaxy density
  profiles}},
\newblock \bibinfo{journal}{Astron. Astrophys.} \bibinfo{volume}{643}
  (\bibinfo{year}{2020}) \bibinfo{pages}{A165}.
  \DOIprefix\doi{10.1051/0004-6361/202038861}.
  \href{http://arxiv.org/abs/2007.02941}{{\tt arXiv:2007.02941}}.
%Type = Article
\bibitem[{Suyu et~al.(2010)Suyu, Marshall, Auger, Hilbert, Blandford, Koopmans,
  Fassnacht, and Treu}]{Suyu:2009by}
\bibinfo{author}{S.~Suyu}, \bibinfo{author}{P.~Marshall},
  \bibinfo{author}{M.~Auger}, \bibinfo{author}{S.~Hilbert},
  \bibinfo{author}{R.~Blandford}, \bibinfo{author}{L.~Koopmans},
  \bibinfo{author}{C.~Fassnacht}, \bibinfo{author}{T.~Treu},
\newblock \bibinfo{title}{{Dissecting the Gravitational Lens B1608+656. II.
  Precision Measurements of the Hubble Constant, Spatial Curvature, and the
  Dark Energy Equation of State}},
\newblock \bibinfo{journal}{Astrophys. J.} \bibinfo{volume}{711}
  (\bibinfo{year}{2010}) \bibinfo{pages}{201--221}.
  \DOIprefix\doi{10.1088/0004-637X/711/1/201}.
  \href{http://arxiv.org/abs/0910.2773}{{\tt arXiv:0910.2773}}.
%Type = Article
\bibitem[{Paraficz and Hjorth(2009)}]{Paraficz:2009xj}
\bibinfo{author}{D.~Paraficz}, \bibinfo{author}{J.~Hjorth},
\newblock \bibinfo{title}{{Gravitational lenses as cosmic rulers: density of
  dark matter and dark energy from time delays and velocity dispersions}},
\newblock \bibinfo{journal}{Astron. Astrophys.} \bibinfo{volume}{507}
  (\bibinfo{year}{2009}) \bibinfo{pages}{L49}.
  \DOIprefix\doi{10.1051/0004-6361/200913307}.
  \href{http://arxiv.org/abs/0910.5823}{{\tt arXiv:0910.5823}}.
%Type = Article
\bibitem[{Jee et~al.(2015)Jee, Komatsu, and Suyu}]{Jee:2014uxa}
\bibinfo{author}{I.~Jee}, \bibinfo{author}{E.~Komatsu}, \bibinfo{author}{S.~H.
  Suyu},
\newblock \bibinfo{title}{{Measuring angular diameter distances of strong
  gravitational lenses}},
\newblock \bibinfo{journal}{JCAP} \bibinfo{volume}{11} (\bibinfo{year}{2015})
  \bibinfo{pages}{033}. \DOIprefix\doi{10.1088/1475-7516/2015/11/033}.
  \href{http://arxiv.org/abs/1410.7770}{{\tt arXiv:1410.7770}}.
%Type = Article
\bibitem[{Suyu et~al.(2017)}]{Suyu:2016qxx}
\bibinfo{author}{S.~Suyu}, et~al.,
\newblock \bibinfo{title}{{H0LiCOW -- I. H0 Lenses in COSMOGRAIL's Wellspring:
  program overview}},
\newblock \bibinfo{journal}{Mon. Not. Roy. Astron. Soc.} \bibinfo{volume}{468}
  (\bibinfo{year}{2017}) \bibinfo{pages}{2590--2604}.
  \DOIprefix\doi{10.1093/mnras/stx483}.
  \href{http://arxiv.org/abs/1607.00017}{{\tt arXiv:1607.00017}}.
%Type = Inproceedings
\bibitem[{Treu et~al.(2013)}]{Treu:2013rpx}
\bibinfo{author}{T.~Treu}, et~al.,
\newblock \bibinfo{title}{{Dark Energy with Gravitational Lens Time Delays}},
\newblock in: \bibinfo{booktitle}{{Community Summer Study 2013}: {Snowmass on
  the Mississippi}}, \bibinfo{year}{2013}.
  \href{http://arxiv.org/abs/1306.1272}{{\tt arXiv:1306.1272}}.
%Type = Article
\bibitem[{Liao et~al.(2015)}]{Liao:2014cka}
\bibinfo{author}{K.~Liao}, et~al.,
\newblock \bibinfo{title}{{Strong Lens Time Delay Challenge: II. Results of
  TDC1}},
\newblock \bibinfo{journal}{Astrophys. J.} \bibinfo{volume}{800}
  (\bibinfo{year}{2015}) \bibinfo{pages}{11}.
  \DOIprefix\doi{10.1088/0004-637X/800/1/11}.
  \href{http://arxiv.org/abs/1409.1254}{{\tt arXiv:1409.1254}}.
%Type = Article
\bibitem[{Jee et~al.(2016)Jee, Komatsu, Suyu, and Huterer}]{Jee:2015yra}
\bibinfo{author}{I.~Jee}, \bibinfo{author}{E.~Komatsu}, \bibinfo{author}{S.~H.
  Suyu}, \bibinfo{author}{D.~Huterer},
\newblock \bibinfo{title}{{Time-delay Cosmography: Increased Leverage with
  Angular Diameter Distances}},
\newblock \bibinfo{journal}{JCAP} \bibinfo{volume}{04} (\bibinfo{year}{2016})
  \bibinfo{pages}{031}. \DOIprefix\doi{10.1088/1475-7516/2016/04/031}.
  \href{http://arxiv.org/abs/1509.03310}{{\tt arXiv:1509.03310}}.
%Type = Article
\bibitem[{Wong et~al.(2017)}]{Wong:2016dpo}
\bibinfo{author}{K.~C. Wong}, et~al.,
\newblock \bibinfo{title}{{H0LiCOW -- IV. Lens mass model of HE 0435$-$1223 and
  blind measurement of its time-delay distance for cosmology}},
\newblock \bibinfo{journal}{Mon. Not. Roy. Astron. Soc.} \bibinfo{volume}{465}
  (\bibinfo{year}{2017}) \bibinfo{pages}{4895--4913}.
  \DOIprefix\doi{10.1093/mnras/stw3077}.
  \href{http://arxiv.org/abs/1607.01403}{{\tt arXiv:1607.01403}}.
%Type = Article
\bibitem[{Bonvin et~al.(2017)}]{Bonvin:2016crt}
\bibinfo{author}{V.~Bonvin}, et~al.,
\newblock \bibinfo{title}{{H0LiCOW -- V. New COSMOGRAIL time delays of HE
  0435$-$1223: $H_0$ to 3.8 per cent precision from strong lensing in a flat
  $\Lambda$CDM model}},
\newblock \bibinfo{journal}{Mon. Not. Roy. Astron. Soc.} \bibinfo{volume}{465}
  (\bibinfo{year}{2017}) \bibinfo{pages}{4914--4930}.
  \DOIprefix\doi{10.1093/mnras/stw3006}.
  \href{http://arxiv.org/abs/1607.01790}{{\tt arXiv:1607.01790}}.
%Type = Article
\bibitem[{Tihhonova et~al.(2018)}]{Tihhonova:2017mym}
\bibinfo{author}{O.~Tihhonova}, et~al.,
\newblock \bibinfo{title}{{H0LiCOW VIII. A weak-lensing measurement of the
  external convergence in the field of the lensed quasar HE 0435$-$1223}},
\newblock \bibinfo{journal}{Mon. Not. Roy. Astron. Soc.} \bibinfo{volume}{477}
  (\bibinfo{year}{2018}) \bibinfo{pages}{5657--5669}.
  \DOIprefix\doi{10.1093/mnras/sty1040}.
  \href{http://arxiv.org/abs/1711.08804}{{\tt arXiv:1711.08804}}.
%Type = Article
\bibitem[{Birrer et~al.(2019)}]{Birrer:2018vtm}
\bibinfo{author}{S.~Birrer}, et~al.,
\newblock \bibinfo{title}{{H0LiCOW - IX. Cosmographic analysis of the doubly
  imaged quasar SDSS 1206+4332 and a new measurement of the Hubble constant}},
\newblock \bibinfo{journal}{Mon. Not. Roy. Astron. Soc.} \bibinfo{volume}{484}
  (\bibinfo{year}{2019}) \bibinfo{pages}{4726}.
  \DOIprefix\doi{10.1093/mnras/stz200}.
  \href{http://arxiv.org/abs/1809.01274}{{\tt arXiv:1809.01274}}.
%Type = Article
\bibitem[{Rusu et~al.(2019)}]{Rusu:2019xrq}
\bibinfo{author}{C.~E. Rusu}, et~al.,
\newblock \bibinfo{title}{{H0LiCOW XII. Lens mass model of WFI2033-4723 and
  blind measurement of its time-delay distance and $H_0$}},
\newblock \bibinfo{journal}{MNRAS}  (\bibinfo{year}{2019}).
  \DOIprefix\doi{10.1093/mnras/stz3451}.
  \href{http://arxiv.org/abs/1905.09338}{{\tt arXiv:1905.09338}}.
%Type = Article
\bibitem[{Wong et~al.(2019)}]{Wong:2019kwg}
\bibinfo{author}{K.~C. Wong}, et~al.,
\newblock \bibinfo{title}{{H0LiCOW XIII. A 2.4\% measurement of $H_{0}$ from
  lensed quasars: $5.3\sigma$ tension between early and late-Universe probes}},
\newblock \bibinfo{journal}{MNRAS}  (\bibinfo{year}{2019}).
  \DOIprefix\doi{10.1093/mnras/stz3094}.
  \href{http://arxiv.org/abs/1907.04869}{{\tt arXiv:1907.04869}}.
%Type = Article
\bibitem[{Chen et~al.(2019)}]{Chen:2019ejq}
\bibinfo{author}{G.~C.-F. Chen}, et~al.,
\newblock \bibinfo{title}{{A SHARP view of H0LiCOW: $H_{0}$ from three
  time-delay gravitational lens systems with adaptive optics imaging}},
\newblock \bibinfo{journal}{Mon. Not. Roy. Astron. Soc.} \bibinfo{volume}{490}
  (\bibinfo{year}{2019}) \bibinfo{pages}{1743--1773}.
  \DOIprefix\doi{10.1093/mnras/stz2547}.
  \href{http://arxiv.org/abs/1907.02533}{{\tt arXiv:1907.02533}}.
%Type = Article
\bibitem[{Refsdal(1964)}]{Refsdal1964b}
\bibinfo{author}{S.~Refsdal},
\newblock \bibinfo{title}{{ On the Possibility of Determining Hubble's
  Parameter and the Masses of Galaxies from the Gravitational Lens Effect}},
\newblock \bibinfo{journal}{Monthly Notices of the Royal Astronomical Society}
  \bibinfo{volume}{128} (\bibinfo{year}{1964}) \bibinfo{pages}{307--310}.
  \URLprefix \url{https://doi.org/10.1093/mnras/128.4.307}.
  \DOIprefix\doi{10.1093/mnras/128.4.307}.
%Type = Article
\bibitem[{Kelly et~al.(2015)}]{Kelly:2014mwa}
\bibinfo{author}{P.~L. Kelly}, et~al.,
\newblock \bibinfo{title}{{Multiple Images of a Highly Magnified Supernova
  Formed by an Early-Type Cluster Galaxy Lens}},
\newblock \bibinfo{journal}{Science} \bibinfo{volume}{347}
  (\bibinfo{year}{2015}) \bibinfo{pages}{1123}.
  \DOIprefix\doi{10.1126/science.aaa3350}.
  \href{http://arxiv.org/abs/1411.6009}{{\tt arXiv:1411.6009}}.
%Type = Article
\bibitem[{Goobar et~al.(2017)}]{Goobar:2016uuf}
\bibinfo{author}{A.~Goobar}, et~al.,
\newblock \bibinfo{title}{{iPTF16geu: A multiply imaged, gravitationally lensed
  type Ia supernova}},
\newblock \bibinfo{journal}{Science} \bibinfo{volume}{356}
  (\bibinfo{year}{2017}) \bibinfo{pages}{291--295}.
  \DOIprefix\doi{10.1126/science.aal2729}.
  \href{http://arxiv.org/abs/1611.00014}{{\tt arXiv:1611.00014}}.
%Type = Article
\bibitem[{Pierel and Rodney(2019)}]{Pierel:2019pnr}
\bibinfo{author}{J.~R. Pierel}, \bibinfo{author}{S.~A. Rodney},
\newblock \bibinfo{title}{{Turning Gravitationally Lensed Supernovae into
  Cosmological Probes}},
\newblock \bibinfo{journal}{Astrophys. J.} \bibinfo{volume}{876}
  (\bibinfo{year}{2019}) \bibinfo{pages}{107}.
  \DOIprefix\doi{10.3847/1538-4357/ab164a}.
  \href{http://arxiv.org/abs/1902.01260}{{\tt arXiv:1902.01260}}.
%Type = Article
\bibitem[{{LSST Science Collaboration} et~al.(2009){LSST Science
  Collaboration}, {Abell}, {Allison}, {Anderson}, {Andrew}
  et~al.}]{LSSTscience}
\bibinfo{author}{{LSST Science Collaboration}}, \bibinfo{author}{P.~A.
  {Abell}}, \bibinfo{author}{J.~{Allison}}, \bibinfo{author}{S.~F. {Anderson}},
  \bibinfo{author}{J.~R. {Andrew}}, et~al.,
\newblock \bibinfo{title}{{LSST Science Book, Version 2.0}},
\newblock \bibinfo{journal}{arXiv e-prints}  (\bibinfo{year}{2009})
  \bibinfo{pages}{arXiv:0912.0201}. \href{http://arxiv.org/abs/0912.0201}{{\tt
  arXiv:0912.0201}}.
%Type = Article
\bibitem[{Marshall et~al.(2017)}]{Marshall:2017wph}
\bibinfo{author}{P.~Marshall}, et~al. (\bibinfo{collaboration}{LSST}),
\newblock \bibinfo{title}{{Science-Driven Optimization of the LSST Observing
  Strategy}}  (\bibinfo{year}{2017}). \DOIprefix\doi{10.5281/zenodo.842713}.
  \href{http://arxiv.org/abs/1708.04058}{{\tt arXiv:1708.04058}}.
%Type = Article
\bibitem[{Goldstein et~al.(2019)Goldstein, Nugent, and
  Goobar}]{Goldstein:2018bue}
\bibinfo{author}{D.~A. Goldstein}, \bibinfo{author}{P.~E. Nugent},
  \bibinfo{author}{A.~Goobar},
\newblock \bibinfo{title}{{Rates and Properties of Supernovae Strongly
  Gravitationally Lensed by Elliptical Galaxies in Time-domain Imaging
  Surveys}},
\newblock \bibinfo{journal}{Astrophys. J. Suppl.} \bibinfo{volume}{243}
  (\bibinfo{year}{2019}) \bibinfo{pages}{6}.
  \DOIprefix\doi{10.3847/1538-4365/ab1fe0}.
  \href{http://arxiv.org/abs/1809.10147}{{\tt arXiv:1809.10147}}.
%Type = Article
\bibitem[{Huber et~al.(2019)}]{Huber:2019ljb}
\bibinfo{author}{S.~Huber}, et~al. (\bibinfo{collaboration}{LSST Dark Energy
  Science}),
\newblock \bibinfo{title}{{Strongly lensed SNe Ia in the era of LSST: observing
  cadence for lens discoveries and time-delay measurements}},
\newblock \bibinfo{journal}{Astron. Astrophys.} \bibinfo{volume}{631}
  (\bibinfo{year}{2019}) \bibinfo{pages}{A161}.
  \DOIprefix\doi{10.1051/0004-6361/201935370}.
  \href{http://arxiv.org/abs/1903.00510}{{\tt arXiv:1903.00510}}.
%Type = Article
\bibitem[{Oguri and Kawano(2003)}]{Oguri:2002ku}
\bibinfo{author}{M.~Oguri}, \bibinfo{author}{Y.~Kawano},
\newblock \bibinfo{title}{{Gravitational lens time delays for distant
  supernovae: break the degeneracy between radial mass profiles and the hubble
  constant}},
\newblock \bibinfo{journal}{Mon. Not. Roy. Astron. Soc.} \bibinfo{volume}{338}
  (\bibinfo{year}{2003}) \bibinfo{pages}{L25--L29}.
  \DOIprefix\doi{10.1046/j.1365-8711.2003.06290.x}.
  \href{http://arxiv.org/abs/astro-ph/0211499}{{\tt arXiv:astro-ph/0211499}}.
%Type = Article
\bibitem[{Yahalomi et~al.(2017)Yahalomi, Schechter, and
  Wambsganss}]{Yahalomi:2017ihe}
\bibinfo{author}{D.~A. Yahalomi}, \bibinfo{author}{P.~L. Schechter},
  \bibinfo{author}{J.~Wambsganss},
\newblock \bibinfo{title}{{A Quadruply Lensed SN Ia: Gaining a
  Time-Delay...Losing a Standard Candle}},
\newblock \bibinfo{journal}{MIT Journal of Undergraduate Research}
  (\bibinfo{year}{2017}). \href{http://arxiv.org/abs/1711.07919}{{\tt
  arXiv:1711.07919}}.
%Type = Article
\bibitem[{Foxley-Marrable et~al.(2018)Foxley-Marrable, Collett, Vernardos,
  Goldstein, and Bacon}]{Foxley-Marrable:2018dzu}
\bibinfo{author}{M.~Foxley-Marrable}, \bibinfo{author}{T.~E. Collett},
  \bibinfo{author}{G.~Vernardos}, \bibinfo{author}{D.~A. Goldstein},
  \bibinfo{author}{D.~Bacon},
\newblock \bibinfo{title}{{The impact of microlensing on the standardization of
  strongly lensed Type Ia supernovae}},
\newblock \bibinfo{journal}{Mon. Not. Roy. Astron. Soc.} \bibinfo{volume}{478}
  (\bibinfo{year}{2018}) \bibinfo{pages}{5081--5090}.
  \DOIprefix\doi{10.1093/mnras/sty1346}.
  \href{http://arxiv.org/abs/1802.07738}{{\tt arXiv:1802.07738}}.
%Type = Article
\bibitem[{Bonvin et~al.(2019)Bonvin, Tihhonova, Millon, Chan, Savary, Huber,
  and Courbin}]{Bonvin:2018lgh}
\bibinfo{author}{V.~Bonvin}, \bibinfo{author}{O.~Tihhonova},
  \bibinfo{author}{M.~Millon}, \bibinfo{author}{J.~Chan},
  \bibinfo{author}{E.~Savary}, \bibinfo{author}{S.~Huber},
  \bibinfo{author}{F.~Courbin},
\newblock \bibinfo{title}{{Impact of the 3D source geometry on time-delay
  measurements of lensed type-Ia Supernovae}},
\newblock \bibinfo{journal}{Astron. Astrophys.} \bibinfo{volume}{621}
  (\bibinfo{year}{2019}) \bibinfo{pages}{A55}.
  \DOIprefix\doi{10.1051/0004-6361/201833405}.
  \href{http://arxiv.org/abs/1805.04525}{{\tt arXiv:1805.04525}}.
%Type = Article
\bibitem[{Holanda et~al.(2016)Holanda, Busti, and Alcaniz}]{Holanda:2015zpz}
\bibinfo{author}{R.~Holanda}, \bibinfo{author}{V.~Busti},
  \bibinfo{author}{J.~Alcaniz},
\newblock \bibinfo{title}{{Probing the cosmic distance duality with strong
  gravitational lensing and supernovae Ia data}},
\newblock \bibinfo{journal}{JCAP} \bibinfo{volume}{02} (\bibinfo{year}{2016})
  \bibinfo{pages}{054}. \DOIprefix\doi{10.1088/1475-7516/2016/02/054}.
  \href{http://arxiv.org/abs/1512.02486}{{\tt arXiv:1512.02486}}.
%Type = Article
\bibitem[{Holanda et~al.(2017)Holanda, Busti, Lima, and
  Alcaniz}]{Holanda:2016msr}
\bibinfo{author}{R.~Holanda}, \bibinfo{author}{V.~Busti},
  \bibinfo{author}{F.~Lima}, \bibinfo{author}{J.~Alcaniz},
\newblock \bibinfo{title}{{Probing the distance-duality relation with high-$z$
  data}},
\newblock \bibinfo{journal}{JCAP} \bibinfo{volume}{09} (\bibinfo{year}{2017})
  \bibinfo{pages}{039}. \DOIprefix\doi{10.1088/1475-7516/2017/09/039}.
  \href{http://arxiv.org/abs/1611.09426}{{\tt arXiv:1611.09426}}.
%Type = Article
\bibitem[{Rana et~al.(2017)Rana, Jain, Mahajan, Mukherjee, and
  Holanda}]{Rana:2017sfr}
\bibinfo{author}{A.~Rana}, \bibinfo{author}{D.~Jain},
  \bibinfo{author}{S.~Mahajan}, \bibinfo{author}{A.~Mukherjee},
  \bibinfo{author}{R.~Holanda},
\newblock \bibinfo{title}{{Probing the cosmic distance duality relation using
  time delay lenses}},
\newblock \bibinfo{journal}{JCAP} \bibinfo{volume}{07} (\bibinfo{year}{2017})
  \bibinfo{pages}{010}. \DOIprefix\doi{10.1088/1475-7516/2017/07/010}.
  \href{http://arxiv.org/abs/1705.04549}{{\tt arXiv:1705.04549}}.
%Type = Article
\bibitem[{Etherington(1933)}]{Etherington}
\bibinfo{author}{I.~M.~H. Etherington},
\newblock \bibinfo{title}{The definition of distance in general relativity},
\newblock \bibinfo{journal}{Philos. Mag.} \bibinfo{volume}{15}
  (\bibinfo{year}{1933}) \bibinfo{pages}{761--773}.
  \DOIprefix\doi{10.1080/14786443309462220}.
%Type = Article
\bibitem[{{Ellis}(2007)}]{Ellis2007}
\bibinfo{author}{G.~F.~R. {Ellis}},
\newblock \bibinfo{title}{{On the definition of distance in general relativity:
  I. M. H. Etherington (Philosophical Magazine ser. 7, vol. 15, 761 (1933))}},
\newblock \bibinfo{journal}{General Relativity and Gravitation}
  \bibinfo{volume}{39} (\bibinfo{year}{2007}) \bibinfo{pages}{1047--1052}.
  \DOIprefix\doi{10.1007/s10714-006-0355-5}.
%Type = Article
\bibitem[{Hehl et~al.(1976)Hehl, Von Der~Heyde, Kerlick, and
  Nester}]{Hehl:1976kj}
\bibinfo{author}{F.~Hehl}, \bibinfo{author}{P.~Von Der~Heyde},
  \bibinfo{author}{G.~Kerlick}, \bibinfo{author}{J.~Nester},
\newblock \bibinfo{title}{{General Relativity with Spin and Torsion:
  Foundations and Prospects}},
\newblock \bibinfo{journal}{Rev. Mod. Phys.} \bibinfo{volume}{48}
  (\bibinfo{year}{1976}) \bibinfo{pages}{393--416}.
  \DOIprefix\doi{10.1103/RevModPhys.48.393}.
%Type = Article
\bibitem[{Hehl et~al.(1995)Hehl, McCrea, Mielke, and Ne'eman}]{Hehl:1994ue}
\bibinfo{author}{F.~W. Hehl}, \bibinfo{author}{J.~McCrea},
  \bibinfo{author}{E.~W. Mielke}, \bibinfo{author}{Y.~Ne'eman},
\newblock \bibinfo{title}{{Metric affine gauge theory of gravity: Field
  equations, Noether identities, world spinors, and breaking of dilation
  invariance}},
\newblock \bibinfo{journal}{Phys. Rept.} \bibinfo{volume}{258}
  (\bibinfo{year}{1995}) \bibinfo{pages}{1--171}.
  \DOIprefix\doi{10.1016/0370-1573(94)00111-F}.
  \href{http://arxiv.org/abs/gr-qc/9402012}{{\tt arXiv:gr-qc/9402012}}.
%Type = Article
\bibitem[{Hammond(2002)}]{Hammond_2002}
\bibinfo{author}{R.~T. Hammond},
\newblock \bibinfo{title}{Torsion gravity},
\newblock \bibinfo{journal}{Reports on Progress in Physics}
  \bibinfo{volume}{65} (\bibinfo{year}{2002}) \bibinfo{pages}{599--649}.
  \URLprefix \url{https://doi.org/10.1088\%2F0034-4885%2F65%2F5%2F201}.
  \DOIprefix\doi{10.1088/0034-4885/65/5/201}.
%Type = Article
\bibitem[{Gabrielli et~al.(2006)Gabrielli, Huitu, and Roy}]{Gabrielli:2006im}
\bibinfo{author}{E.~Gabrielli}, \bibinfo{author}{K.~Huitu},
  \bibinfo{author}{S.~Roy},
\newblock \bibinfo{title}{{Photon propagation in magnetic and electric fields
  with scalar/pseudoscalar couplings: A New look}},
\newblock \bibinfo{journal}{Phys. Rev. D} \bibinfo{volume}{74}
  (\bibinfo{year}{2006}) \bibinfo{pages}{073002}.
  \DOIprefix\doi{10.1103/PhysRevD.74.073002}.
  \href{http://arxiv.org/abs/hep-ph/0604143}{{\tt arXiv:hep-ph/0604143}}.
%Type = Article
\bibitem[{Santana et~al.(2017)Santana, Calvão, Reis, and
  Siffert}]{Santana:2017zvy}
\bibinfo{author}{L.~T. Santana}, \bibinfo{author}{M.~O. Calvão},
  \bibinfo{author}{R.~R.~R. Reis}, \bibinfo{author}{B.~B. Siffert},
\newblock \bibinfo{title}{{How does light move in a generic metric-affine
  background?}},
\newblock \bibinfo{journal}{Phys. Rev. D} \bibinfo{volume}{95}
  (\bibinfo{year}{2017}) \bibinfo{pages}{061501}.
  \DOIprefix\doi{10.1103/PhysRevD.95.061501}.
  \href{http://arxiv.org/abs/1703.10871}{{\tt arXiv:1703.10871}}.
%Type = Article
\bibitem[{Bassett and Kunz(2004)}]{Bassett:2003vu}
\bibinfo{author}{B.~A. Bassett}, \bibinfo{author}{M.~Kunz},
\newblock \bibinfo{title}{{Cosmic distance-duality as a probe of exotic physics
  and acceleration}},
\newblock \bibinfo{journal}{Phys. Rev. D} \bibinfo{volume}{69}
  (\bibinfo{year}{2004}) \bibinfo{pages}{101305}.
  \DOIprefix\doi{10.1103/PhysRevD.69.101305}.
  \href{http://arxiv.org/abs/astro-ph/0312443}{{\tt arXiv:astro-ph/0312443}}.
%Type = Article
\bibitem[{Avgoustidis et~al.(2009)Avgoustidis, Verde, and
  Jimenez}]{Avgoustidis2009}
\bibinfo{author}{A.~Avgoustidis}, \bibinfo{author}{L.~Verde},
  \bibinfo{author}{R.~Jimenez},
\newblock \bibinfo{title}{{Consistency among distance measurements:
  transparency, {BAO} scale and accelerated expansion}},
\newblock \bibinfo{journal}{J. Cosmol. Astropart. Phys.} \bibinfo{volume}{2009}
  (\bibinfo{year}{2009}). \DOIprefix\doi{10.1088/1475-7516/2009/06/012}.
  \href{http://arxiv.org/abs/0902.2006}{{\tt arXiv:0902.2006}}.
%Type = Article
\bibitem[{Avgoustidis et~al.(2010)Avgoustidis, Burrage, Redondo, Verde, and
  Jimenez}]{Avgoustidis2010}
\bibinfo{author}{A.~Avgoustidis}, \bibinfo{author}{C.~Burrage},
  \bibinfo{author}{J.~Redondo}, \bibinfo{author}{L.~Verde},
  \bibinfo{author}{R.~Jimenez},
\newblock \bibinfo{title}{Constraints on cosmic opacity and beyond the standard
  model physics from cosmological distance measurements},
\newblock \bibinfo{journal}{J. Cosmol. Astropart. Phys.} \bibinfo{volume}{2010}
  (\bibinfo{year}{2010}). \DOIprefix\doi{10.1088/1475-7516/2010/10/024}.
  \href{http://arxiv.org/abs/1004.2053}{{\tt arXiv:1004.2053}}.
%Type = Article
\bibitem[{Astier et~al.(2014)}]{Astier:2014swa}
\bibinfo{author}{P.~Astier}, et~al.,
\newblock \bibinfo{title}{{Extending the supernova Hubble diagram to z $\sim$
  1.5 with the Euclid space mission}},
\newblock \bibinfo{journal}{Astron. Astrophys.} \bibinfo{volume}{572}
  (\bibinfo{year}{2014}) \bibinfo{pages}{A80}.
  \DOIprefix\doi{10.1051/0004-6361/201423551}.
  \href{http://arxiv.org/abs/1409.8562}{{\tt arXiv:1409.8562}}.
%Type = Article
\bibitem[{Coe and Moustakas(2009)}]{Coe_2009}
\bibinfo{author}{D.~Coe}, \bibinfo{author}{L.~A. Moustakas},
\newblock \bibinfo{title}{Cosmological constraints from gravitational lens time
  delays},
\newblock \bibinfo{journal}{The Astrophysical Journal} \bibinfo{volume}{706}
  (\bibinfo{year}{2009}) \bibinfo{pages}{45–59}. \URLprefix
  \url{http://dx.doi.org/10.1088/0004-637X/706/1/45}.
  \DOIprefix\doi{10.1088/0004-637x/706/1/45}.
%Type = Article
\bibitem[{Linder(2011)}]{Linder:2011dr}
\bibinfo{author}{E.~V. Linder},
\newblock \bibinfo{title}{{Lensing Time Delays and Cosmological
  Complementarity}},
\newblock \bibinfo{journal}{Phys. Rev. D} \bibinfo{volume}{84}
  (\bibinfo{year}{2011}) \bibinfo{pages}{123529}.
  \DOIprefix\doi{10.1103/PhysRevD.84.123529}.
  \href{http://arxiv.org/abs/1109.2592}{{\tt arXiv:1109.2592}}.
%Type = Article
\bibitem[{{Torrado} and {Lewis}(2020)}]{Torrado:2020dgo}
\bibinfo{author}{J.~{Torrado}}, \bibinfo{author}{A.~{Lewis}},
\newblock \bibinfo{title}{{Cobaya: Code for Bayesian Analysis of hierarchical
  physical models}},
\newblock \bibinfo{journal}{arXiv e-prints}  (\bibinfo{year}{2020})
  \bibinfo{pages}{arXiv:2005.05290}.
  \href{http://arxiv.org/abs/2005.05290}{{\tt arXiv:2005.05290}}.
%Type = Article
\bibitem[{Avgoustidis et~al.(2010)Avgoustidis, Burrage, Redondo, Verde, and
  Jimenez}]{Avgoustidis:2010ju}
\bibinfo{author}{A.~Avgoustidis}, \bibinfo{author}{C.~Burrage},
  \bibinfo{author}{J.~Redondo}, \bibinfo{author}{L.~Verde},
  \bibinfo{author}{R.~Jimenez},
\newblock \bibinfo{title}{{Constraints on cosmic opacity and beyond the
  standard model physics from cosmological distance measurements}},
\newblock \bibinfo{journal}{JCAP} \bibinfo{volume}{1010} (\bibinfo{year}{2010})
  \bibinfo{pages}{024}. \DOIprefix\doi{10.1088/1475-7516/2010/10/024}.
  \href{http://arxiv.org/abs/1004.2053}{{\tt arXiv:1004.2053}}.
%Type = Article
\bibitem[{Martinelli et~al.(2020)}]{Martinelli:2020hud}
\bibinfo{author}{M.~Martinelli}, et~al. (\bibinfo{collaboration}{EUCLID}),
\newblock \bibinfo{title}{{Euclid: Forecast constraints on the cosmic distance
  duality relation with complementary external probes}},
\newblock \bibinfo{journal}{Astronomy \& Astrophysics}  (\bibinfo{year}{2020}).
  \href{http://arxiv.org/abs/2007.16153}{{\tt arXiv:2007.16153}}.
%Type = Article
\bibitem[{Hogg et~al.(2020)Hogg, Martinelli, and Nesseris}]{Hogg:2020ktc}
\bibinfo{author}{N.~B. Hogg}, \bibinfo{author}{M.~Martinelli},
  \bibinfo{author}{S.~Nesseris},
\newblock \bibinfo{title}{{Constraints on the distance duality relation with
  standard sirens}},
\newblock \bibinfo{journal}{JCAP} \bibinfo{volume}{12} (\bibinfo{year}{2020})
  \bibinfo{pages}{019}. \DOIprefix\doi{10.1088/1475-7516/2020/12/019}.
  \href{http://arxiv.org/abs/2007.14335}{{\tt arXiv:2007.14335}}.
%Type = Article
\bibitem[{Bogdanos and Nesseris(2009)}]{Bogdanos:2009ib}
\bibinfo{author}{C.~Bogdanos}, \bibinfo{author}{S.~Nesseris},
\newblock \bibinfo{title}{{Genetic Algorithms and Supernovae Type Ia
  Analysis}},
\newblock \bibinfo{journal}{JCAP} \bibinfo{volume}{05} (\bibinfo{year}{2009})
  \bibinfo{pages}{006}. \DOIprefix\doi{10.1088/1475-7516/2009/05/006}.
  \href{http://arxiv.org/abs/0903.2805}{{\tt arXiv:0903.2805}}.
%Type = Article
\bibitem[{Nesseris and Garcia-Bellido(2012)}]{Nesseris:2012tt}
\bibinfo{author}{S.~Nesseris}, \bibinfo{author}{J.~Garcia-Bellido},
\newblock \bibinfo{title}{{A new perspective on Dark Energy modeling via
  Genetic Algorithms}},
\newblock \bibinfo{journal}{JCAP} \bibinfo{volume}{1211} (\bibinfo{year}{2012})
  \bibinfo{pages}{033}. \DOIprefix\doi{10.1088/1475-7516/2012/11/033}.
  \href{http://arxiv.org/abs/1205.0364}{{\tt arXiv:1205.0364}}.
%Type = Article
\bibitem[{Nesseris and Shafieloo(2010)}]{Nesseris:2010ep}
\bibinfo{author}{S.~Nesseris}, \bibinfo{author}{A.~Shafieloo},
\newblock \bibinfo{title}{{A model independent null test on the cosmological
  constant}},
\newblock \bibinfo{journal}{Mon. Not. Roy. Astron. Soc.} \bibinfo{volume}{408}
  (\bibinfo{year}{2010}) \bibinfo{pages}{1879--1885}.
  \DOIprefix\doi{10.1111/j.1365-2966.2010.17254.x}.
  \href{http://arxiv.org/abs/1004.0960}{{\tt arXiv:1004.0960}}.
%Type = Article
\bibitem[{Nesseris and Garcia-Bellido(2013)}]{Nesseris:2013bia}
\bibinfo{author}{S.~Nesseris}, \bibinfo{author}{J.~Garcia-Bellido},
\newblock \bibinfo{title}{{Comparative analysis of model-independent methods
  for exploring the nature of dark energy}},
\newblock \bibinfo{journal}{Phys. Rev.} \bibinfo{volume}{D88}
  (\bibinfo{year}{2013}) \bibinfo{pages}{063521}.
  \DOIprefix\doi{10.1103/PhysRevD.88.063521}.
  \href{http://arxiv.org/abs/1306.4885}{{\tt arXiv:1306.4885}}.
%Type = Article
\bibitem[{Sapone et~al.(2014)Sapone, Majerotto, and Nesseris}]{Sapone:2014nna}
\bibinfo{author}{D.~Sapone}, \bibinfo{author}{E.~Majerotto},
  \bibinfo{author}{S.~Nesseris},
\newblock \bibinfo{title}{{Curvature versus distances: Testing the FLRW
  cosmology}},
\newblock \bibinfo{journal}{Phys. Rev. D} \bibinfo{volume}{90}
  (\bibinfo{year}{2014}) \bibinfo{pages}{023012}.
  \DOIprefix\doi{10.1103/PhysRevD.90.023012}.
  \href{http://arxiv.org/abs/1402.2236}{{\tt arXiv:1402.2236}}.
%Type = Article
\bibitem[{{Arjona} and {Nesseris}(2019)}]{Arjona:2019fwb}
\bibinfo{author}{R.~{Arjona}}, \bibinfo{author}{S.~{Nesseris}},
\newblock \bibinfo{title}{{What can Machine Learning tell us about the
  background expansion of the Universe?}},
\newblock \bibinfo{journal}{arXiv e-prints}  (\bibinfo{year}{2019})
  \bibinfo{pages}{arXiv:1910.01529}.
  \href{http://arxiv.org/abs/1910.01529}{{\tt arXiv:1910.01529}}.
%Type = Article
\bibitem[{{Arjona} and {Nesseris}(2020)}]{Arjona:2020kco}
\bibinfo{author}{R.~{Arjona}}, \bibinfo{author}{S.~{Nesseris}},
\newblock \bibinfo{title}{{Hints of dark energy anisotropic stress using
  Machine Learning}},
\newblock \bibinfo{journal}{arXiv e-prints}  (\bibinfo{year}{2020})
  \bibinfo{pages}{arXiv:2001.11420}.
  \href{http://arxiv.org/abs/2001.11420}{{\tt arXiv:2001.11420}}.
%Type = Article
\bibitem[{Arjona(2020)}]{Arjona:2020doi}
\bibinfo{author}{R.~Arjona},
\newblock \bibinfo{title}{{Machine Learning meets the redshift evolution of the
  CMB Temperature}},
\newblock \bibinfo{journal}{JCAP}  (\bibinfo{year}{2020}).
  \href{http://arxiv.org/abs/2002.12700}{{\tt arXiv:2002.12700}}.
%Type = Book
\bibitem[{Rasmussen and Williams(2006)}]{Rasmussen}
\bibinfo{author}{C.~E. Rasmussen}, \bibinfo{author}{C.~K.~I. Williams},
  \bibinfo{title}{Gaussian Processes for Machine Learning},
  \bibinfo{publisher}{MIT Press}, \bibinfo{year}{2006}.
%Type = Article
\bibitem[{{Holsclaw} et~al.(2010{\natexlab{a}}){Holsclaw}, {Alam}, {Sans{\'o}},
  {Lee}, {Heitmann}, {Habib}, and {Higdon}}]{Holsclaw2010a}
\bibinfo{author}{T.~{Holsclaw}}, \bibinfo{author}{U.~{Alam}},
  \bibinfo{author}{B.~{Sans{\'o}}}, \bibinfo{author}{H.~{Lee}},
  \bibinfo{author}{K.~{Heitmann}}, \bibinfo{author}{S.~{Habib}},
  \bibinfo{author}{D.~{Higdon}},
\newblock \bibinfo{title}{{Nonparametric reconstruction of the dark energy
  equation of state}},
\newblock \bibinfo{journal}{Physical Review D} \bibinfo{volume}{82}
  (\bibinfo{year}{2010}{\natexlab{a}}) \bibinfo{pages}{103502}.
  \DOIprefix\doi{10.1103/PhysRevD.82.103502}.
  \href{http://arxiv.org/abs/1009.5443}{{\tt arXiv:1009.5443}}.
%Type = Article
\bibitem[{{Holsclaw} et~al.(2010{\natexlab{b}}){Holsclaw}, {Alam}, {Sans{\'o}},
  {Lee}, {Heitmann}, {Habib}, and {Higdon}}]{Holsclaw2010b}
\bibinfo{author}{T.~{Holsclaw}}, \bibinfo{author}{U.~{Alam}},
  \bibinfo{author}{B.~{Sans{\'o}}}, \bibinfo{author}{H.~{Lee}},
  \bibinfo{author}{K.~{Heitmann}}, \bibinfo{author}{S.~{Habib}},
  \bibinfo{author}{D.~{Higdon}},
\newblock \bibinfo{title}{{Nonparametric Dark Energy Reconstruction from
  Supernova Data}},
\newblock \bibinfo{journal}{Physical Review Letters} \bibinfo{volume}{105}
  (\bibinfo{year}{2010}{\natexlab{b}}) \bibinfo{pages}{241302}.
  \DOIprefix\doi{10.1103/PhysRevLett.105.241302}.
  \href{http://arxiv.org/abs/1011.3079}{{\tt arXiv:1011.3079}}.
%Type = Article
\bibitem[{{Holsclaw} et~al.(2011){Holsclaw}, {Alam}, {Sans{\'o}}, {Lee},
  {Heitmann}, {Habib}, and {Higdon}}]{Holsclaw2011}
\bibinfo{author}{T.~{Holsclaw}}, \bibinfo{author}{U.~{Alam}},
  \bibinfo{author}{B.~{Sans{\'o}}}, \bibinfo{author}{H.~{Lee}},
  \bibinfo{author}{K.~{Heitmann}}, \bibinfo{author}{S.~{Habib}},
  \bibinfo{author}{D.~{Higdon}},
\newblock \bibinfo{title}{{Nonparametric reconstruction of the dark energy
  equation of state from diverse data sets}},
\newblock \bibinfo{journal}{Physical Review Letters} \bibinfo{volume}{84}
  (\bibinfo{year}{2011}) \bibinfo{pages}{083501}.
  \DOIprefix\doi{10.1103/PhysRevD.84.083501}.
  \href{http://arxiv.org/abs/1104.2041}{{\tt arXiv:1104.2041}}.
%Type = Article
\bibitem[{{Shafieloo} et~al.(2012){Shafieloo}, {Kim}, and
  {Linder}}]{Shafieloo:2012ht}
\bibinfo{author}{A.~{Shafieloo}}, \bibinfo{author}{A.~G. {Kim}},
  \bibinfo{author}{E.~V. {Linder}},
\newblock \bibinfo{title}{{Gaussian process cosmography}},
\newblock \bibinfo{journal}{Physical Review D} \bibinfo{volume}{85}
  (\bibinfo{year}{2012}) \bibinfo{pages}{123530}.
  \DOIprefix\doi{10.1103/PhysRevD.85.123530}.
  \href{http://arxiv.org/abs/1204.2272}{{\tt arXiv:1204.2272}}.
%Type = Article
\bibitem[{{Seikel} et~al.(2012){Seikel}, {Clarkson}, and {Smith}}]{Seikel2012}
\bibinfo{author}{M.~{Seikel}}, \bibinfo{author}{C.~{Clarkson}},
  \bibinfo{author}{M.~{Smith}},
\newblock \bibinfo{title}{{Reconstruction of dark energy and expansion dynamics
  using Gaussian processes}},
\newblock \bibinfo{journal}{Journal of Cosmology and Astroparticle Physics}
  \bibinfo{volume}{6} (\bibinfo{year}{2012}) \bibinfo{pages}{036}.
  \DOIprefix\doi{10.1088/1475-7516/2012/06/036}.
  \href{http://arxiv.org/abs/1204.2832}{{\tt arXiv:1204.2832}}.
%Type = Article
\bibitem[{{Zhang} and {Li}(2018)}]{Zhang2018}
\bibinfo{author}{M.-J. {Zhang}}, \bibinfo{author}{H.~{Li}},
\newblock \bibinfo{title}{{Gaussian processes reconstruction of dark energy
  from observational data}},
\newblock \bibinfo{journal}{European Physical Journal C} \bibinfo{volume}{78}
  (\bibinfo{year}{2018}) \bibinfo{pages}{460}.
  \DOIprefix\doi{10.1140/epjc/s10052-018-5953-3}.
  \href{http://arxiv.org/abs/1806.02981}{{\tt arXiv:1806.02981}}.
%Type = Article
\bibitem[{{Martinelli} et~al.(2019){Martinelli}, {Hogg}, {Peirone}, {Bruni},
  and {Wands}}]{Martinelli:2019dau}
\bibinfo{author}{M.~{Martinelli}}, \bibinfo{author}{N.~B. {Hogg}},
  \bibinfo{author}{S.~{Peirone}}, \bibinfo{author}{M.~{Bruni}},
  \bibinfo{author}{D.~{Wands}},
\newblock \bibinfo{title}{{Constraints on the interacting vacuum -- geodesic
  CDM scenario}},
\newblock \bibinfo{journal}{Monthly Notices of the Royal Astronomical Society}
  \bibinfo{volume}{488} (\bibinfo{year}{2019}) \bibinfo{pages}{3423--3438}.
  \DOIprefix\doi{10.1093/mnras/stz1915}.
  \href{http://arxiv.org/abs/1902.10694}{{\tt arXiv:1902.10694}}.
%Type = Article
\bibitem[{Gerardi et~al.(2019)Gerardi, Martinelli, and
  Silvestri}]{Gerardi:2019obr}
\bibinfo{author}{F.~Gerardi}, \bibinfo{author}{M.~Martinelli},
  \bibinfo{author}{A.~Silvestri},
\newblock \bibinfo{title}{{Reconstruction of the Dark Energy equation of state
  from latest data: the impact of theoretical priors}},
\newblock \bibinfo{journal}{JCAP} \bibinfo{volume}{1907} (\bibinfo{year}{2019})
  \bibinfo{pages}{042}. \DOIprefix\doi{10.1088/1475-7516/2019/07/042}.
  \href{http://arxiv.org/abs/1902.09423}{{\tt arXiv:1902.09423}}.
%Type = Article
\bibitem[{Hogg et~al.(2020)Hogg, Bruni, Crittenden, Martinelli, and
  Peirone}]{Hogg:2020rdp}
\bibinfo{author}{N.~B. Hogg}, \bibinfo{author}{M.~Bruni},
  \bibinfo{author}{R.~Crittenden}, \bibinfo{author}{M.~Martinelli},
  \bibinfo{author}{S.~Peirone},
\newblock \bibinfo{title}{{Latest evidence for a late time vacuum -- geodesic
  CDM interaction}},
\newblock \bibinfo{journal}{Phys. Dark Univ.} \bibinfo{volume}{29}
  (\bibinfo{year}{2020}) \bibinfo{pages}{100583}.
  \DOIprefix\doi{10.1016/j.dark.2020.100583}.
  \href{http://arxiv.org/abs/2002.10449}{{\tt arXiv:2002.10449}}.
%Type = Article
\bibitem[{Seikel and Clarkson(2013)}]{Seikel:2013fda}
\bibinfo{author}{M.~Seikel}, \bibinfo{author}{C.~Clarkson},
\newblock \bibinfo{title}{{Optimising Gaussian processes for reconstructing
  dark energy dynamics from supernovae}}  (\bibinfo{year}{2013}).
  \href{http://arxiv.org/abs/1311.6678}{{\tt arXiv:1311.6678}}.
%Type = Phdthesis
\bibitem[{Duvenaud(2014)}]{Duvenaud2014}
\bibinfo{author}{D.~Duvenaud}, \bibinfo{title}{Automatic model construction
  with Gaussian processes}, Ph.D. thesis, University of Cambridge,
  \bibinfo{year}{2014}.
%Type = Article
\bibitem[{Pedregosa et~al.(2011)Pedregosa, Varoquaux, Gramfort, Michel,
  Thirion, Grisel, Blondel, Prettenhofer, Weiss, Dubourg, Vanderplas, Passos,
  Cournapeau, Brucher, Perrot, and Duchesnay}]{scikit-learn}
\bibinfo{author}{F.~Pedregosa}, \bibinfo{author}{G.~Varoquaux},
  \bibinfo{author}{A.~Gramfort}, \bibinfo{author}{V.~Michel},
  \bibinfo{author}{B.~Thirion}, \bibinfo{author}{O.~Grisel},
  \bibinfo{author}{M.~Blondel}, \bibinfo{author}{P.~Prettenhofer},
  \bibinfo{author}{R.~Weiss}, \bibinfo{author}{V.~Dubourg},
  \bibinfo{author}{J.~Vanderplas}, \bibinfo{author}{A.~Passos},
  \bibinfo{author}{D.~Cournapeau}, \bibinfo{author}{M.~Brucher},
  \bibinfo{author}{M.~Perrot}, \bibinfo{author}{E.~Duchesnay},
\newblock \bibinfo{title}{Scikit-learn: Machine learning in {P}ython},
\newblock \bibinfo{journal}{Journal of Machine Learning Research}
  \bibinfo{volume}{12} (\bibinfo{year}{2011}) \bibinfo{pages}{2825--2830}.
%Type = Article
\bibitem[{Lewis(2019)}]{GetDist}
\bibinfo{author}{A.~Lewis},
\newblock \bibinfo{title}{{GetDist: a Python package for analysing Monte Carlo
  samples}}  (\bibinfo{year}{2019}).
  \href{http://arxiv.org/abs/1910.13970}{{\tt arXiv:1910.13970}}.
%Type = Article
\bibitem[{Hunter(2007)}]{Matplotlib}
\bibinfo{author}{J.~D. Hunter},
\newblock \bibinfo{title}{Matplotlib: A 2d graphics environment},
\newblock \bibinfo{journal}{Computing in Science \& Engineering}
  \bibinfo{volume}{9} (\bibinfo{year}{2007}) \bibinfo{pages}{90--95}.
  \DOIprefix\doi{10.1109/MCSE.2007.55}.
%Type = Article
\bibitem[{{Harris} et~al.(2020){Harris}, {Jarrod Millman}, {van der Walt},
  {Gommers}, {Virtanen}, {Cournapeau}, {Wieser}, {Taylor}, {Berg}, {Smith},
  {Kern}, {Picus}, {Hoyer}, {van Kerkwijk}, {Brett}, {Haldane}, {Fern{\'a}ndez
  del R{\'\i}o}, {Wiebe}, {Peterson}, {G{\'e}rard-Marchant}, {Sheppard},
  {Reddy}, {Weckesser}, {Abbasi}, {Gohlke}, and {Oliphant}}]{NumPy}
\bibinfo{author}{C.~R. {Harris}}, \bibinfo{author}{K.~{Jarrod Millman}},
  \bibinfo{author}{S.~J. {van der Walt}}, \bibinfo{author}{R.~{Gommers}},
  \bibinfo{author}{P.~{Virtanen}}, \bibinfo{author}{D.~{Cournapeau}},
  \bibinfo{author}{E.~{Wieser}}, \bibinfo{author}{J.~{Taylor}},
  \bibinfo{author}{S.~{Berg}}, \bibinfo{author}{N.~J. {Smith}},
  \bibinfo{author}{R.~{Kern}}, \bibinfo{author}{M.~{Picus}},
  \bibinfo{author}{S.~{Hoyer}}, \bibinfo{author}{M.~H. {van Kerkwijk}},
  \bibinfo{author}{M.~{Brett}}, \bibinfo{author}{A.~{Haldane}},
  \bibinfo{author}{J.~{Fern{\'a}ndez del R{\'\i}o}},
  \bibinfo{author}{M.~{Wiebe}}, \bibinfo{author}{P.~{Peterson}},
  \bibinfo{author}{P.~{G{\'e}rard-Marchant}}, \bibinfo{author}{K.~{Sheppard}},
  \bibinfo{author}{T.~{Reddy}}, \bibinfo{author}{W.~{Weckesser}},
  \bibinfo{author}{H.~{Abbasi}}, \bibinfo{author}{C.~{Gohlke}},
  \bibinfo{author}{T.~E. {Oliphant}},
\newblock \bibinfo{title}{{Array Programming with NumPy}},
\newblock \bibinfo{journal}{Nature} \bibinfo{volume}{585}
  (\bibinfo{year}{2020}) \bibinfo{pages}{357–362}.
%Type = Article
\bibitem[{{Virtanen} et~al.(2020){Virtanen}, {Gommers}, {Oliphant},
  {Haberland}, {Reddy}, {Cournapeau}, {Burovski}, {Peterson}, {Weckesser},
  {Bright}, {van der Walt}, {Brett}, {Wilson}, {Jarrod Millman}, {Mayorov},
  {Nelson}, {Jones}, {Kern}, {Larson}, {Carey}, {Polat}, {Feng}, {Moore}, {Vand
  erPlas}, {Laxalde}, {Perktold}, {Cimrman}, {Henriksen}, {Quintero}, {Harris},
  {Archibald}, {Ribeiro}, {Pedregosa}, {van Mulbregt}, and
  {Contributors}}]{2020SciPy-NMeth}
\bibinfo{author}{P.~{Virtanen}}, \bibinfo{author}{R.~{Gommers}},
  \bibinfo{author}{T.~E. {Oliphant}}, \bibinfo{author}{M.~{Haberland}},
  \bibinfo{author}{T.~{Reddy}}, \bibinfo{author}{D.~{Cournapeau}},
  \bibinfo{author}{E.~{Burovski}}, \bibinfo{author}{P.~{Peterson}},
  \bibinfo{author}{W.~{Weckesser}}, \bibinfo{author}{J.~{Bright}},
  \bibinfo{author}{S.~J. {van der Walt}}, \bibinfo{author}{M.~{Brett}},
  \bibinfo{author}{J.~{Wilson}}, \bibinfo{author}{K.~{Jarrod Millman}},
  \bibinfo{author}{N.~{Mayorov}}, \bibinfo{author}{A.~R.~J. {Nelson}},
  \bibinfo{author}{E.~{Jones}}, \bibinfo{author}{R.~{Kern}},
  \bibinfo{author}{E.~{Larson}}, \bibinfo{author}{C.~{Carey}},
  \bibinfo{author}{{\.I}.~{Polat}}, \bibinfo{author}{Y.~{Feng}},
  \bibinfo{author}{E.~W. {Moore}}, \bibinfo{author}{J.~{Vand erPlas}},
  \bibinfo{author}{D.~{Laxalde}}, \bibinfo{author}{J.~{Perktold}},
  \bibinfo{author}{R.~{Cimrman}}, \bibinfo{author}{I.~{Henriksen}},
  \bibinfo{author}{E.~A. {Quintero}}, \bibinfo{author}{C.~R. {Harris}},
  \bibinfo{author}{A.~M. {Archibald}}, \bibinfo{author}{A.~H. {Ribeiro}},
  \bibinfo{author}{F.~{Pedregosa}}, \bibinfo{author}{P.~{van Mulbregt}},
  \bibinfo{author}{S.~.~. {Contributors}},
\newblock \bibinfo{title}{{SciPy 1.0: Fundamental Algorithms for Scientific
  Computing in Python}},
\newblock \bibinfo{journal}{Nature Methods} \bibinfo{volume}{17}
  (\bibinfo{year}{2020}) \bibinfo{pages}{261--272}.
  \DOIprefix\doi{https://doi.org/10.1038/s41592-019-0686-2}.

\end{thebibliography}

\end{document}